# Predictable markets? A news-driven model of the stock market

Maxim Gusev[1], Dimitri Kroujiline*[2], Boris Govorkov[1], Sergey V. Sharov[3], Dmitry Ushanov[4] and Maxim Zhilyaev[5]

[1] IBC Quantitative Strategies, Tärnaby, Sweden, www.ibctrends.com.

[2] LGT Capital Partners, Pfäffikon, Switzerland, www.lgt-capital-partners.com. Email: dimitri.kroujiline@lgt.com. (*corresponding author)

[3] N.I. Lobachevsky State University, Advanced School of General & Applied Physics, Nizhny Novgorod, Russia, www.unn.ru.

[4] Moscow State University, Department of Mechanics and Mathematics, Moscow, Russia, www.math.msu.su.

[5] Mozilla Corporation, Mountain View, CA, USA, www.mozilla.org.

**Abstract**: We attempt to explain stock market dynamics in terms of the interaction among three variables: market price, investor opinion and information flow. We propose a framework for such interaction and apply it to build a model of stock market dynamics which we study both empirically and theoretically. We demonstrate that this model replicates observed market behavior on all relevant timescales (from days to years) reasonably well. Using the model, we obtain and discuss a number of results that pose implications for current market theory and offer potential practical applications.

Keywords: stock market, market dynamics, return predictability, news analysis, language patterns, investor behavior, herding, business cycle, sentiment evolution, reference sentiment level, volatility, return distribution, Ising, agent-based models, price feedback, nonlinear dynamical systems.

# Introduction

There is a simple chain of events that leads to price changes. Prices change when investors buy or sell securities and it is the flow of information that influences the opinions of investors, according to which they make investment decisions. Although this description is too general to be of practical use, it highlights the point that with the correct choice of hypotheses about how prices, opinions and information interact it could be possible to model market dynamics.

In the first part of this paper (Section 1), we develop a mechanism that links information to opinions and opinions to prices. First, we select the market for US stocks as the object of study due to its depth, breadth and high standing in the financial community, which places it in the focus of global financial news media resulting in the dissemination of large amounts of relevant information. We then collect information using proprietary online news aggregators as well as news archives offered by data providers. Next we analyze collected news with respect to their influence on investors' market views. Our approach is to treat each individual news item as if it were a "sales pitch" that motivates investors to either enter into or withdraw from the market and apply methods from marketing research to assess effectiveness.

We then proceed to model opinion dynamics. It is reasonable to expect that, along with the impact of information, such a model should also account for interactions among investors and various idiosyncratic factors that can be assumed to act as random disturbances. There are similarities between this problem and some well-studied problems in statistical mechanics, allowing us to borrow from the existing toolkit to derive an equation for the evolution of investor opinion in analytic form. At this juncture, we investigate the connection between opinions and prices. To understand it, we must answer questions pertaining to how investors make decisions; for instance, is an investor more likely to invest if her market outlook has been stably positive or whether it has recently improved? We suggest a simple solution based on observed investor behavior.

The procedure described above is used to construct a time series of daily prices from 1996 to 2012 based on collected information and compare it with the time series of actual market prices from that same period. We demonstrate that the model replicates the observed market behavior over the studied period reasonably well. In the end of Section 1 we report and discuss results that may be relevant for market theory and practical application.

Although this (empirical) model enables us to translate a given information flow into market prices, it cannot sufficiently highlight the nature of complex market behaviors precisely for the reason that information is treated in the model as a given. To gain further insight into the origins of market dynamics we must extend this framework by including information as a variable, along with investor opinion and market price. In the second part of this paper (Section 2), we incorporate assumptions on how information can be generated and channeled throughout the market to develop a closed-form, self-contained model of stock market dynamics for theoretical study.

We find that information supplied to the market can be represented by two components that play important but different roles in market dynamics. The first component, which consists of news caused by price changes themselves, induces a feedback loop whereby information impacts price and price impacts information. The resulting nonlinear dynamic explains the appearance in the model of essential elements of market behavior such as trends, rallies and crashes and leads to the familiar non-normal shape of the return distribution. Additionally, this dynamic entails a complex causal relation between information and price, as cause and effect become, in a sense, intertwined.

The second informational component contains any other relevant news. It acts as a random external force that drives market dynamics away from equilibrium and, from time to time, triggers changes of market regimes.

We conclude Section 2 by comparing the characteristic behaviors of theoretically-modeled, empirically-modeled and observed market prices and by discussing the possibility of market forecasts. Finally, we provide an overall summary of conclusions in Section 3.

The present study has been carried out with a view toward potentially predicting stock market returns. This view is supported by the empirical research over the last 30 years, which suggests that returns are predictable, especially over long horizons (see Fama and French (1988, 1989), Campbell and Shiller (1988a,b), Cochrane (1999, 2008), Baker and Wurgler (2000), Campbell and Thompson (2008)). This empirical research has primarily focused on identifying potentially predictive variables, such as the dividend yield, earnings-price ratio, credit spread and others, and verifying or refuting their correlation with subsequent returns, typically applying regression methods.

Our objective is to capture the basic mechanisms underlying this predictability.[1] We show that the theoretical model (Section 2), which treats information, opinion and price as endogenous variables, can reproduce observed market features reasonably well, including the price path and the return distribution, under the realistic choice of parameter values consistent with the values obtained using the empirical data (Section 1). Most importantly, this model permits market regimes where deterministic dynamics dominate random behaviors, implying that returns are, in principle, predictable. Because this model is fundamentally nonlinear, the causal relation among the variables is substantially more complex than regression dependence. It therefore follows that the standard approach to return prediction, based on regression methods, may need to be redefined. Such an approach would combine, analogous to weather forecasting, theoretical models with empirical data to predict returns (see Section 2.2.3).

---

[1] The existing literature examines the long-term predictability (e.g. monthly and annual returns) and links it to economic fluctuations and changes in risk perception of investors. Our findings support the view that long-term predictability exists and, furthermore, indicate that returns may already be predictable on intervals of several days. We reach this conclusion using a framework which attributes price changes to the dynamics of investor opinion. Interestingly, we find that, over long horizons, there is a connection between the evolution of investor opinion and economic fluctuations (Fig. 9, Section 1.4.1).

The stock market model developed in this paper belongs to the family of Ising-type models from statistical mechanics (see Section 1.2.1.). The Ising model is a subset of agent-based models – theoretical models that attempt to explain macroscopic phenomena based on the behaviors of individual agents. When applied to economics, the Ising model simulates dynamics among interacting agents capable of making discrete decisions (e.g. buy, sell or hold) subject to random fluctuations (due to idiosyncratic disturbances) and, if any, external influence (e.g. publicly available information) and various constraints (e.g. wealth optimization). As there are numerous choices for factors affecting the dynamics of agents, the main goal is the selection of a reasonably simple combination that allows the replication of distinctive features of actual market behavior in a model. A number of agent-based models have been proposed in the context of financial markets, e.g. Levy, Levy and Solomon (1994, 1995), Caldarelli, Marsili and Zhang (1997), Lux and Marchesi (1999, 2000), Cont and Bouchaud (2000), Sornette and Zhou (2006), Zhou and Sornette (2007). These models differ by choices made in the selection of factors, such as, for example, the form of agent heterogeneity, the topology of interaction, the number of decision-making choices, the form of the agents' utility functions and the nature of external influence on agents.

In the present model we chose to apply the minimal number of possible opinions (buy or sell) and the simplest interaction pattern (all-to-all) to facilitate its study. We also made no a priori assumptions on investors' behaviors, preferences or trading strategies. Our contribution is focused on other areas. First, we identify informational patterns that can effectively influence investors' opinions and measure the corresponding information flow (Section 1.1). Second, we analyze opinion dynamics in the presence of this information flow using a classic (homogeneous) Ising model and construct an empirical time series describing the evolution of investor opinion (Section 1.2). Third, we deduce the relation between investor opinion and market price. While in the existing literature opinions are typically equated with investment decisions, resulting in price changes being proportional to the difference between the number of positive opinions (buyers) and negative opinions (sellers) in a model [2], we derive an equation for price formation by proposing that investors tend to act on their opinions differently over short and long horizons and use this equation to obtain an empirical time series of model prices (Section 1.3). Fourth, we formulate a theoretical model of market dynamics as a heterogeneous Ising model (Section 2) which consists of two types of agents: investors (whose function is, naturally, to invest and divest) and "analysts" (whose function is to interpret news, form opinions and channel them to investors). This model yields a closed-form, nonlinear dynamical system shown to generate behaviors that are in agreement with both the empirical model and the actual market.

This paper, which is primarily intended for economists and investment professionals, is based on ideas and methods from various scientific and practical disciplines, including statistical mechanics and dynamical systems. To preserve its readability, we have endeavored to strike a balance between the depth of the material and the ease of its presentation. To this end, we have placed technical discussions and derivations in the appendices and provided the first principles-based explanations of utilized concepts to make the work self-contained.

## 1. Part I – Empirical study of stock market dynamics

In the empirical study we develop a model that translates information into opinions and opinions into prices. Section 1.1 examines which information can be relevant in the financial markets context and outlines our approach to measuring it and constructing the empirical time series. Section 1.2 derives a model of opinion dynamics and applies it to the time series from the previous section. Section 1.3 develops a model of price formation that, based on the modeled investor opinion, produces the empirical prices for comparison with the actual stock market prices. Section 1.4 discusses results and applications. The relevant technical details are in Appendices A and B.

---

[2] Sometimes price changes are assumed to be proportional to a function of the difference between the number of positive and negative opinions, e.g. Zhang (1999) applied a square root of this difference.

## 1.1. Information

Information comes in many forms. It differs across various dimensions such as content, source, pattern and distribution channel, and there are myriad possibilities for "slicing and dicing" it. Choices made on how information is handled may lead to different theories and practical applications.

Different methods of handling information form the foundations of quintessentially different trading strategies in the field. For example, systematic hedge fund managers apply quantitative techniques to analyze price data to detect trends; global macro managers base their bets on macroeconomic factors; and fundamental long/short funds employ financial analysis to identify mispriced assets.[3]

---

[3] We note that increases in the quantity of information on the internet and its accessibility, the popularity of social networks and improvements in natural language processing and statistical machine learning during the last decade have prompted the development of trading strategies where news analytics complement traditional sources of information. This has also motivated recent empirical research on the correlation between disseminated information, including social media content, and price movements (e.g. Schumaker and Chen (2009); Li et al. (2011); Rechenthin, Street and Srinivasan (2013); Preis, Moat and Stanley (2013)). This paper broadly shares the motivation and aims to advance research in this area by employing a novel approach.

In the theory of finance, a seminal hypothesis on information's influence on investors – the efficient market hypothesis – was developed by Fama (1965, 1970) and Samuelson (1965). The efficient market hypothesis essentially considers that, as a result of competition among rational investors, information related to past events, as well as to anticipated future events, is reflected in spot prices. This theory implies that prices change only as unexpected new information is received and, because such information is random, price changes must also be random variables that cannot be predicted.

We base our approach on the observation that in the context of financial markets information is important only due to its impact on investors. Therefore, it would be sensible to consider only those informational patterns that can effectively impact investors' decision making. We put forward a hypothesis that the effectiveness of information is determined by the degree of directness of its interpretation in relation to the expectations of future market performance, and assume that information which is more direct will be more effectively perceived by investors.

We will test this hypothesis on empirical data in the next sections. Here we briefly discuss the rationale behind it. The basic idea is intuitively simple: we view each news item as if it were a "sales pitch" to buy or sell the market aimed at investors. Sales professionals will confirm that a straightforward, unambiguous message is imperative for it to be effective, and in the present case the least ambiguous message is the one that tells investors directly whether the market is expected to go up or down.

Information that does not succinctly spell this message out requires individually subjective interpretation leading to conflicting views as to its implications for anticipated market performance. We may suppose that news which cannot be easily interpreted in terms of market reaction will result in a roughly equal number of positive and negative views or in a lack of strong opinions altogether and so in average are unlikely to impact market prices. Conversely, news that can be

Page **8** of **97**

easily interpreted will be quickly interpreted (or will have already been interpreted prior to any news release as a market scenario) and will then be followed by information about the expected market reaction. It therefore seems reasonable to conclude that information suggesting the direction of the expected market movement is important for price development. The quantifying of such patterns – henceforth referred to as ***direct information*** – in the general news flow is an area of our research focus.

Thus, in the general daily news flow, we wish to measure the number of news releases that contain this direct information. We limit our study to the US stock market and select the S&P 500 Index as its proxy. In particular, we consider only the English language media and search for patterns that indicate future returns, e.g. "the S&P 500 will increase / decrease", or those which imply a trend, e.g. "the S&P 500 has grown / fallen", assuming that investors would react to such information by extrapolating the trend into the future.

We have chosen not to assign weights to news items as a measure of their relative importance. This factor should emerge naturally by capturing the repetitions or echoing of the original release by other news outlets. Thus, to measure the direct information correctly, we collect all relevant news releases[4] including the duplicates.

Based on the above considerations, we have devised a number of rules and applied them to daily news data for the period from 1996 to 2012 retrieved from DJ/Factiva database. These same rules have also been applied to the news that we have directly collected on a daily basis from online news sources since 2010 (one of the advantages of assembling a proprietary news archive is the ability to control the selection of news sources to ensure data consistency over time).

---

[4] We also take into account news reach, which is a marketing research term, referring here to the number of people exposed to a news outlet during a given period, akin to a newspaper's circulation.

We measure the number of news releases containing positive or negative direct information for each day where the NYSE was open for business during 1996-2012 to obtain a time series of daily direct information $H(t)$ over that period:

$$H(t) = \frac{H_+(t) - H_-(t)}{H_T(t)},$$

where $H_+$ is the number of news items containing positive direct information, $H_-$ is the number of news items containing negative direct information and $H_T$ is the number of all relevant news items (e.g. where the phrase "S&P 500" is mentioned); ordinarily, $H_T > H_+ + H_-$ since $H_T$ contains neutral information along with $H_+$ and $H_-$. For illustration, the time series of daily $H(t)$ and $H_T(t)$ are displayed on a sample interval 2010-2012 (Fig. 1).[5]

---

[5] Technical remarks:

1. $H(t)$ from DJ/Factiva for 1996-2012 is used as an input into the market model that we develop over the next sections. In this context it is important to note that the volume $H_T(t)$ gradually rose during that period, fluctuating on average in the range of 25-50 per day in 1996-2001, 50-100 per day in 2002-2009 and 100-200 per day in 2010-2012. Accordingly, we have to exercise caution with regards to interpreting modeling results for the period 1996-2001 due to lower data quality.

2. $H(t)$ from both sources have been calibrated to reduce (but not eliminate) the positive bias of the original series. This has been achieved by increasing the weight of $H_-(t)$ by 30% in the case of DJ/Factiva and 45% in the case of the proprietary data. The calibration is applied to achieve more realistic model behavior; we note, however, that the results obtained with and without the calibration are not substantially different. The excess positive bias may be partially related to general asymmetry in the perception of negative and positive events (Baumeister et al., 2001), which, for example, implies an aversion to losses in the context of financial markets (Kahneman and Tversky, 1979), however this discussion falls outside the scope of the present work.

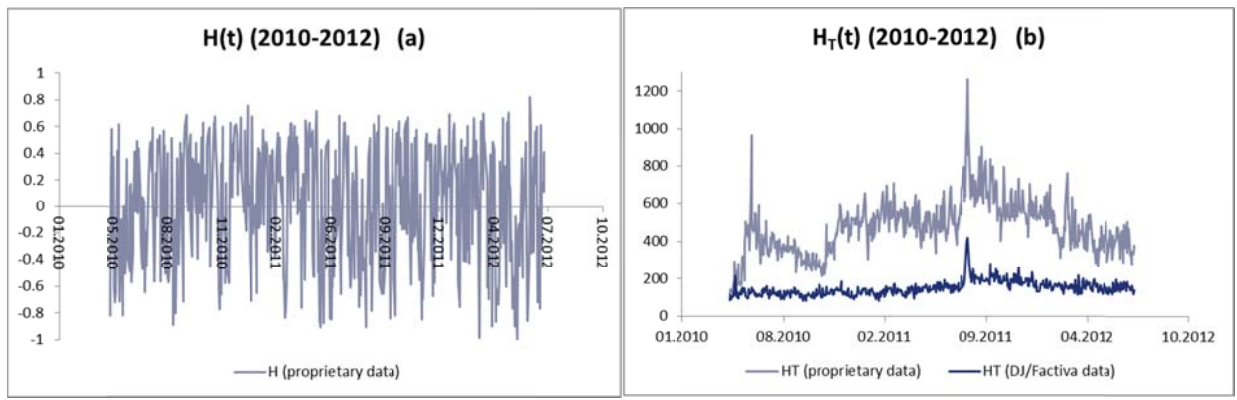

**Figure 1**: (a) Daily $H(t)$ from the proprietary archive (2010-2012). (b) Daily $H_T(t)$ from the proprietary archive and DJ/Factiva archive (2010-2012). During this period, $H_T$ was on average 475 per day based on the proprietary data and 150 per day based on DJ/Factiva data. We note that $H_T$ based on DJ/Factiva data has substantially increased in 2013, becoming comparable with $H_T$ based on the proprietary data.

## 1.2. Sentiment

We define (market) ***sentiment*** of an investor as her summary view on future market performance. We do not differentiate between "rational" and "irrational" behavior as investment decisions are usually driven by a mixture of both the analysis and, consciously or unconsciously, the "gut feeling" (due, in part, to behavioral biases). Thus, sentiment is defined here to comprise all types of opinions, whether formed through rational analysis or irrational beliefs, based on all possible informational inputs.[6]

---

3. To obtain some of the results reported in the next sections, we use zero-phase digital filtering methods that, in our situation, require roughly 300 data points on each end of the time series for calculation. We apply these methods to the extended time series (1995-2013) based on DJ/Factiva data, so that the corresponding results are (strictly speaking) correct only in the interval 1996-2012. For consistency, we report the data, results and conclusions for the period 1996-2012 throughout this paper.

In the previous section we have proposed that direct information can effectively influence investors' views. In this section we use an analogy from physics to explore how investor sentiment can evolve in response to direct information flow.

### 1.2.1. Model of sentiment dynamics

In this section we make a first principles-based introduction to the Ising model (Ising, 1925) from statistical mechanics, as it applies to the problem presented here, and follow with a preliminary discussion of the relevant effects. Although the Ising model has been broadly applied to study

---

[6] We note that the term sentiment appears in the finance literature usually in the sense of an opinion prompted by feelings or beliefs, as opposed to an opinion reached through rational analysis. In this context, the notion of sentiment has been applied to describe the driving forces behind price deviations from fundamental values that cannot be explained within the classic framework of rational decision making. Various empirical measures of sentiment utilized in the literature (e.g. Brown and Cliff (2004), Baker and Wurgler (2007), Lux (2011)) include indices based on periodic surveys of investor opinion as well as the proxies such as closed-end fund discounts, advancing vs. declining issues, call vs. put contracts and others.

Our approach relies on the fact that it is the investor opinion per se that leads to investment decisions, irrespective of whether it has been formed rationally or irrationally. The understanding of how this summary opinion – which we have herein defined as sentiment – evolves is one of the goals of the present work. Accordingly, in this section we model the sentiment dynamics without invoking explicit assumptions on the rationality of agents and apply this model to obtain sentiment empirically from the time series of direct information as measured in Section 1.1.

problems in social and economic dynamics, we have not come across its explanation from first principles in the socioeconomic context.

We consider a model with a large number of investors ($N \gg 1$), identical in all respects except the ability to form differing binary opinions ($\pm 1$) as to whether the market will rise or fall. Let us say that the $i$-th investor has the sentiment $s_i = +1$ if she opines that the market will rise and $s_i = -1$ if she opines that the market will fall. We introduce a function, called energy in physics, that describes the macroscopic states of such a system. This function should contain those factors for which we wish to account in the model. Basically, there are two such factors.

The first factor captures the impact due to the flow of direct information. Earlier we assumed that direct information, either positive (the market will go up) or negative (the market will go down), can be effective in forcing investors to change their opinions, i.e. direct information acts to orient the investors' sentiments along its (positive or negative) direction. We express the energy of impact on the $i$-th investor as $-H_i s_i$, where $H_i$ is direct information received by the investor. Energy is negative where the investor's opinion and direct information are coaligned because $s_i$ and $H_i$ are of the same sign and is positive otherwise. It means that, according to the familiar principle of minimum energy from physics, direct information acts to (re)align sentiment along its direction to reduce the energy. The overall energy due to the impact of direct information flow on the investors is $-\mu \sum_i^N H_i s_i$, obtained by summing $-H_i s_i$ for all investors and where $\mu$ is a positive coefficient determining the strength of the impact.

The second factor is the interaction among the individual investors (agents). When people exchange opinions, they may influence the opinions of others and in turn be influenced as well. So we can generally say that people tend to "coalign" their opinions. In our context, where investors exchange views about market performance (i.e. sentiments), we can write the energy of interaction between the $i$-th and $j$-th investors as $-J_{ij} s_i s_j$, where $J_{ij}$ is a positive coefficient determining the strength of interaction. Energy will be minimal (negative) when the sentiments are coaligned and maximal (positive) otherwise, which means that, in line with the principle of minimum energy, interaction tends to make investors co-orient their views. The overall energy of interaction $-\frac{1}{2}\sum_{i\neq j}^{N} J_{ij} s_i s_j$ is obtained by summing $-J_{ij} s_i s_j$ across all pairs of investors ($i \neq j$) and multiplying the sum by ½ to eliminate double counting.

Consequently, the total energy of the system takes the form:

$$E = -\frac{1}{2}\sum_{i\neq j}^{N} J_{ij} s_i s_j - \mu \sum_{i}^{N} H_i(t) s_i \ . \tag{1}$$

Provided that the same information is available to all investors, it follows that the energy (1) is minimized if investors' sentiments are coaligned. We should therefore expect to find that all investors in the model share the same positive or negative market outlook. Yet we know that the reality is different: there will always be investors whose market sentiments are contrary to the popular opinion. This discrepancy arises for the reason that our model does not integrate the infinite number of different specific influences experienced by investors that lead to randomness and disorder present in real world.

To incorporate the effect of this randomness we again borrow from physics where temperature serves as a measure of disorder – the higher the temperature, the more stochastic-like a system's behavior. We apply this same methodology and introduce an economic analog to temperature, which we henceforth simply call **temperature** or $\theta$, that will indicate the degree of disorder in our model. This means that each investor will be subject to random disturbances that may cause her to occasionally change sentiment irrespective of other investors in the model; in other words, investors' sentiments will be subject to random fluctuations. As a result, the study of system (1) now requires statistical methods of analysis that we will apply in the next section.

Let us briefly discuss the effects that such microscopic random fluctuations may have on the macroscopic characteristics of the system. At low temperatures, random fluctuations are weak, so that the interactions among investors lead the system toward one of two ordered macroscopic, polarized states where the total sentiment is either positive or negative. In this case, according to the model, investors' behavior is characterized by a high degree of herding that continues to increase as the temperature falls. At high temperatures, random influences prevail and the system as a whole appears disordered, with total sentiment fluctuating around zero, i.e. investors fail to establish a consensus opinion and proceed to act randomly. In physics, the above-described macroscopic states are called phases and when a system changes state it is known as phase transition.

In reality it is probably seldom, if at all, that investors behave either randomly or in perfect synchronicity. It would be reasonable to suppose that the actual market is mostly confined to a transitional regime between disordered and ordered phases, where random behavior and herding or crowd behavior can be of roughly equal importance. In Section 1.3.2 we will provide evidence in support of this conjecture.

This family of models, originally developed to describe ferromagnetism in statistical mechanics, is broadly called the Ising model. The Ising model has been applied to many problems in social dynamics over the last 30 years (see Castellano, Fortunato and Loreto (2009)). In economics, the Ising model was utilized for the first time by Vaga (1990), who adapted it to financial markets to infer the existence of certain characteristic market regimes. Since the mid-1990s there has been extensive economic research in this area and the related field of agent-based modeling (see Levy, Levy and Solomon (2000), Samanidou et al. (2007), Lux (2009) and Sornette (2014)).

Our application of the Ising model differs from other research in this area in two main aspects: (i) in the empirical part of the paper (Section 1), we use direct information flow, which we can measure, as an external force in the (homogeneous) model and arrive at a closed-form dynamic equation that governs the evolution of the system's sentiment; and (ii) in the theoretical part of the paper (Section 2), we develop a heterogeneous, two-component extension of the above model to gain insight into the dynamics of the interaction between sentiment and direct information.

### 1.2.2. Equation for sentiment evolution

As noted earlier, system (1), in which investors' sentiments are subject to random fluctuation, necessitates statistical methods of analysis. In this section we study the evolution of the system's total sentiment as a statistical average[7]: $s = \langle s_T \rangle / N$, where $s_T = \sum_i^N s_i$ and $\langle\ \rangle$ denotes the statistical average. We note that $-1 \leq s \leq 1$ because $\langle s_T \rangle$ can vary between $-N$ and $+N$.

To derive the equation for $s(t)$ in analytic form, we make two simplifying assumptions: (i) all investors receive the same information and (ii) each investor interacts with all other investors with the same strength, which yields an all-to-all interaction pattern[8]. As a result, the system's total energy (1) can be written as

$$E = -\frac{1}{2} J_0 \sum_{i \neq j}^N s_i s_j - \mu H(t) \sum_i^N s_i, \tag{2}$$

---

[7] Also called the ensemble average. To distinguish between the statistical or ensemble average, on the one hand, and the average with respect to time, on the other hand, we denote the former using angle brackets and the latter using horizontal bars.

[8] The use of the all-to-all interaction pattern is a sensible first step for studying this problem as it is also the leading-order approximation for a general interaction topology in the Ising model. We will discuss the implications of this approximation in Section 2.2.3.

where $H(t)$ is the uniform information flow $(H_i = H(t))$ and $J_0$ is the constant strength of interaction $(J_{ik} = J_0)$.

A consideration of the kinetics of system (2) leads to the following equation for $s(t)$ (Appendix A, eq. A16):[9]

$$\dot{s} = -w_s s + w_s \tanh\left(\frac{Js + \mu H(t)}{\theta}\right), \tag{3}$$

where the dot denotes the derivative with respect to time, $J = NJ_0$ is the rescaled strength of interaction[10], $\theta$ has been introduced in the previous section as the economic analog to temperature, i.e. the parameter which indicates the level of disorder in the system, and $w_s = 1/\tau_s$ with $\tau_s$ defined as the characteristic time over which random disturbances will make individual sentiment $s_i$ flip and so indicates the investor's average memory time-span.

Let us consider equation (3). To begin with, it would be more convenient to rewrite it as

$$\dot{s} = F(s,t) = -w_s s + w_s \tanh(\beta_1 s + \beta_2 H(t)), \tag{4}$$

where $F$ has the meaning of the total force acting on the sentiment $s$, $\beta_1 = J/\theta$ is a dimensionless parameter which, being inversely proportional to temperature, determines the degree of order in the system and $\beta_2 = \mu/\theta$.

---

[9] Equation (3) is obtained in this paper as a special case of the system of dynamic equations derived in the statistical limit $N \to \infty$ for the all-to-all interaction pattern (see Appendix A) that we study in Section 2. Equation (3) was originally obtained by Suzuki and Kubo (1968), who used the mean-field approximation to derive it.

[10] Note that $J_0 \sim 1/N$ ensures that coupling energy is finite in the all-to-all interaction case, therefore $J = O(1)$.

For illustration purposes let us assume that $\tanh(\beta_2 \sigma_H) \ll 1$, where $\sigma_H$ is the standard deviation of the time series $H(t)$.[11] This assumption enables us to represent $F$ by a sum of the time-independent and time-dependent components, which takes the following leading order form:

$$F(s,t) = -w_s(s - \tanh(\beta_1 s)) + w_s \operatorname{sech}^2(\beta_1 s) \tanh(\beta_2 H(t)).$$

Equation (4) can then be written as

$$m\ddot{s} + \dot{s} = -\frac{dU_0(s)}{ds} + w_s \operatorname{sech}^2(\beta_1 s) \tanh(\beta_2 H(t)), \tag{5}$$

where the coefficient $m$ is zero in accordance with (4) (we will need $m$ for interpreting (5) in the next paragraph) and the time-independent component of $F$ has been expressed via the function $U_0(s)$, called potential, given with the precision up to a constant by

$$U_0(s) = w_s \left(\frac{1}{2}s^2 - \frac{1}{\beta_1}\ln\cosh(\beta_1 s)\right). \tag{6}$$

And so, we have arrived at the equation for an overdamped, forced nonlinear oscillator. The overdamping means that inertia ($m\ddot{s}$) is small and can be neglected relative to damping ($\dot{s}$). Thus equation (5) can be interpreted as governing the motion of a zero-mass ($m = 0$) damped particle driven by the force applied, which is dependent on $H(t)$, inside the potential well, the shape of which is determined by $U_0(s)$. Accordingly, $s$ takes on the meaning of the particle's coordinate. Therefore the motion of the particle will trace the evolution path of the sentiment.

In other words, equation (5) describes a situation where the time-dependent force (information flow) acts to displace the particle (sentiment) inside the potential well from its at-rest equilibrium (where the consensus of opinion is reached) while the restoring force (the interactions among investors subject to random influences) counteracts it by compelling sentiment back toward equilibrium.[12]

---

[11] As we will see later, $\beta_2 \sigma_H < 1$ (Table I), so that this assumption is within the relevant range of parameter values.

The expression for the potential $U_0(s)$ (eq. 6) reveals the existence of disordered and ordered states within the system as a function of temperature.[13] The potential is symmetric. It has the U-shape with one stable equilibrium point $s = 0$ for $\beta_1 < 1$ (the high temperature phase: $\theta > J$) and the W-shape with one unstable ($s = 0$) and two stable ($s = s_\pm$) equilibrium points that are symmetric with respect to the origin for $\beta_1 > 1$ (the low temperature phase: $\theta < J$) (Fig. 2). The U-shape corresponds to the disordered state since sentiment is zero at equilibrium reflecting the fact that investors tend to behave randomly in this phase. The W-shape corresponds to the ordered state where sentiment settles at either the negative ($s_-$) or positive ($s_+$) value at equilibrium, as herding behavior prevails in this phase. Thus, the model contains the phase regimes that were discussed in Section 1.2.1.

Figure 2 shows that in the disordered state ($\beta_1 < 1$) decreasing $\beta_1$ causes the potential well to contract, so that sentiment becomes entrenched around zero. Similarly, in the ordered state ($\beta_1 > 1$) when $\beta_1$ increases the negative well ($s < 0$) and the positive well ($s > 0$) quickly deepen and simultaneously shift toward the boundaries ($s = \pm 1$), so that sentiment becomes trapped at the extreme negative or positive values. It follows that, as we have expected, the transitional regime $\beta_1 \approx 1$ is likely to be prevalent, otherwise the latitude of the change of sentiment would be unrealistically restricted.

---

[12] This motion is finite as $-1 \leq s \leq 1$, $\dot{s} = F < 0$ at $s = 1$ and $\dot{s} = F > 0$ at $s = -1$. However, the system does not permit free oscillations around equilibrium: the absence of inertia makes the particle fall directly toward the equilibrium points, which are the stable nodes in dynamical systems terminology.

[13] It may be easier to understand the behavior of $U_0(s)$ if it is expanded into a truncated Taylor series in the powers of $s$ as $U_0(s) = -w_s \left( \frac{\beta_1 - 1}{2} s^2 - \frac{\beta_1^3}{12} s^4 \right)$, which holds reasonably well for any $s$ when $\beta_1 \sim 1$. Note that the phase transition at $\beta_1 = 1$ corresponds to the change of sign of the quadratic term.

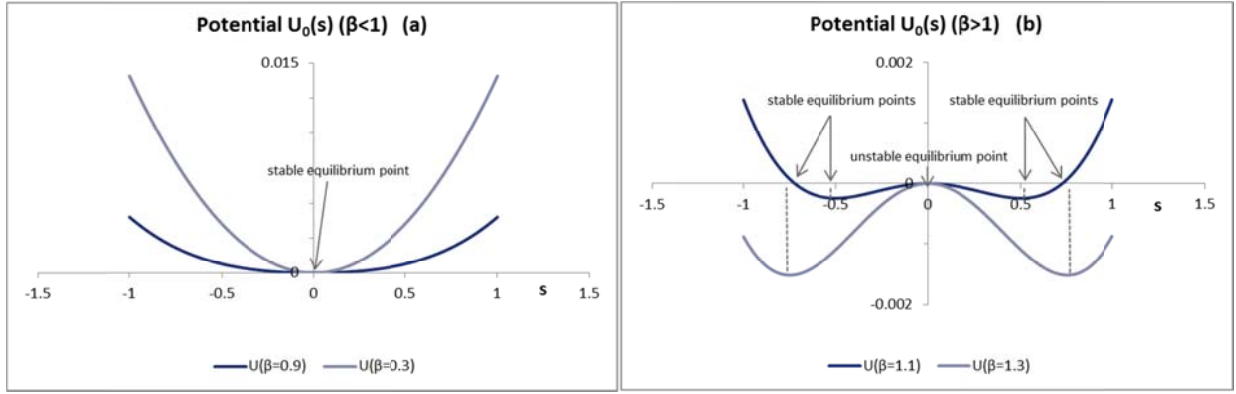

**Figure 2**: (a) The potential $U_0(s)$ for $\beta_1 = 0.3$ and 0.9. (b) $U_0(s)$ for $\beta_1 = 1.1$ and 1.3. The location of the stable equilibrium points corresponds to the minima of $U_0(s)$ and the unstable equilibrium points to the maxima of $U_0(s)$. There is one stable equilibrium point in the disordered state (Fig. 2a) and one unstable and two stable equilibrium points in the ordered state (Fig. 2b).

Finally, it is worthwhile noting that the absence of inertia in equation (5) implies that the model has no intrinsic memory and so it would generally be impossible to predict the evolution of sentiment for an arbitrary $H(t)$. The possibility of making a meaningful forecast is then determined solely by the statistical properties of $H(t)$. For example, it could in principle be possible to predict the change of sentiment if $H(t)$ is autocorrelated. We will return to this point in Section 1.4.2.

### 1.2.3. Modeled sentiment

In the previous two sections we have developed the model that converts direct information $H$ into investors' market sentiment $s$ (eq. 4). Now we can construct the empirical time series $s(t)$ from the daily measured $H(t)$, but for that we must first estimate the values of the parameters $\beta_1$, $\beta_2$ and $w_s$.

First, as noted earlier, it would be reasonable to consider a situation where both random behavior and herding behavior are equally present, which corresponds to $\beta_1 \sim 1$. So we set $\beta_1 = 1$ as the initial estimate. Second, it may prove useful to assign $\beta_2 \sim 1$ so that the term $\beta_2 H$ is neither dominating nor negligible. We therefore take $\beta_2 = 1$ as the initial estimate. Third, intuitively, typical investor's memory horizon lies somewhere in the window from several days to several months. It seems reasonable to choose as the initial guess $\tau_s$ corresponding to 25 business days, which is roughly 1 month. Accordingly we estimate $w_s = 1/\tau_s = 0.04$.

We substitute the daily series $H(t)$ into equation (4) with $\beta_1 = 1, \beta_2 = 1$ and $w_s = 0.04$ to obtain the daily series $s(t)$. Figure 3 shows $s(t)$ and the S&P 500 Index for periods of different length between 1996 and 2012. Our first observation is that there are certain similarities between the behaviors of the modelled sentiment and the actual index. For instance, there appears to be a correspondence between the turning points of sentiment evolution and the turning points of index trends on various timescales. This observation is supported by the statistics that the increments of sentiment values and index log returns are cross-correlated at 55% on a daily basis, 69% on a weekly basis and 75% on a monthly basis at zero time lag.

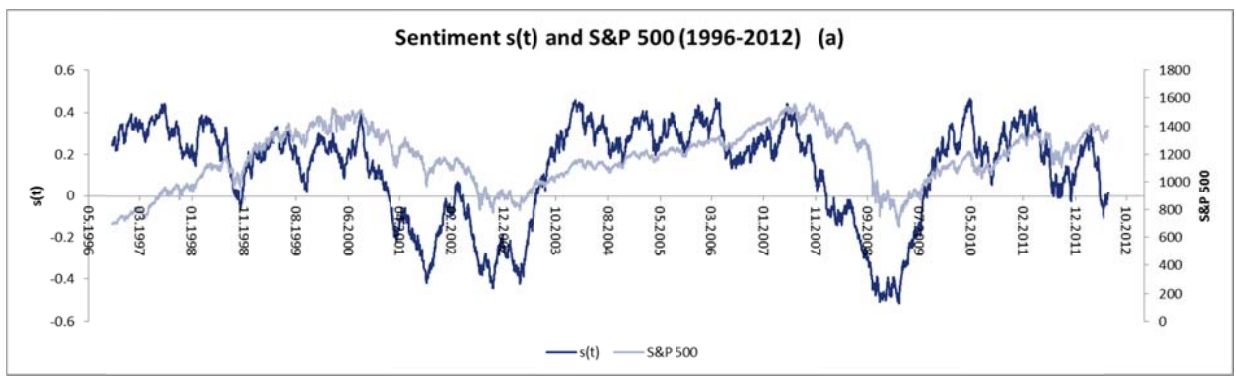

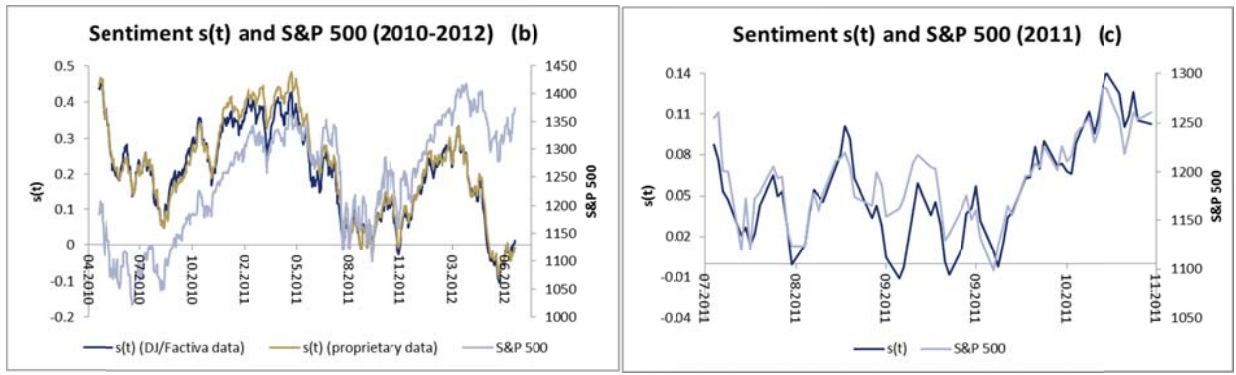

**Figure 3**: The sentiment $s(t)$ constructed from $H(t)$ for $w_s = 0.04, \beta_1 = 1,$ and $\beta_2 = 1.0$ and the S&P 500 Index during (a) 1996-2012; (b) 2010-2012; and (c) 07/2011-11/2011. Note the close similarity between $s(t)$ based on DJ/Factiva data and $s(t)$ based on the proprietary data (Fig. 3b).

Our second observation concerns two distinct behaviors in sentiment: a bounded motion within the positive region between $s = 0.2$ and $s = 0.4$ that corresponds to mid-term index trends (upward and downward) and infrequent large-scale movements into and out of the negative region that correspond to market crashes and rallies. Although we have set $\beta_1 = 1$, such behaviors are more compatible with the W-shape of the potential $U_0(s)$ for $\beta_1 > 1$ (eq. 6) as the former can be related to motion inside the positive well while the latter can be related to motion across the negative and positive wells. We will demonstrate that the W-shape is indeed the prevalent configuration of the potential well in Section 1.3.2.

We can also see that sentiment remains positive most of the time, even during relatively long stretches of market losses, and becomes negative only during pronounced bear markets. This observation is surprising given that we have considered the period during which the US economy experienced the largest financial crisis since the Great Depression and the stock market spent two thirds of this time below its high-water mark and one quarter of this time in bear market return territory. It hints at the possibility that there is a natural tendency for sentiment to be positively skewed, which would be consistent with the asymmetry of the long-term behavior observed in the stock market (perhaps best exemplified by the fact that the DJ Industrial Average Index has returned on average around 7.5% p.a. over the last 100 years).

According to our model, it requires a positive mean of $H(t)$ ($\bar{H} > 0$) to keep sentiment positive on average, i.e. the volume of positive direct information must, in the long run, exceed that of negative direct information. Indeed, the daily mean $\bar{H}$ measured for the period covered by our data is positive (Table I). This positive information bias is possibly related to the economic growth and inflation.

The asymmetry of sentiment's behavior induced by $\bar{H} > 0$ can also be explained in terms of the perturbation of potential well. If we decompose $H$ into two parts as $H(t) = \bar{H} + H'(t)$, where $\bar{H}$ is the constant mean of $H(t)$ and $H'(t)$ is its time-dependent component, such that $\overline{H'(t)} = 0$, then equations (5) and (6) can be written as

$$\dot{s} = -\frac{dU_c(s)}{ds} + w_s \text{sech}^2(\beta_1 s + c) \tanh(\beta_2 H'(t)), \tag{7a}$$

with

$$U_c(s) = w_s \left( \frac{1}{2} s^2 - \frac{1}{\beta_1} \ln \cosh(\beta_1 s + c) \right), \tag{7b}$$

where $c = \beta_2 \bar{H} > 0$.

Figure 4 shows that a positive $c$ breaks the symmetry of the potential. For $\beta_1 > 1$ the positive well ($s > 0$) deepens and the negative well ($s < 0$) flattens, so that the probability of $s$ crossing into the negative territory decreases.[14] For larger $c$ the shallow well is further distorted to the level beyond which the negative stable equilibrium point ($s = s_-$) vanishes and the double-well configuration transforms into a single-well configuration in which the sentiment is positive at the equilibrium. The overall effect of this distortion is that sentiment tends to spend more time in the positive well interrupted by relatively short excursions into negative sentiment territory. The figure shows that even a small $c$ can have a pronounced effect on the shape of the potential well (e.g. $c \gtrsim 0.01$ for $\beta_1 \approx 1$). Using the relevant values for $\beta_2$ and $\bar{H}$ (Table I), we obtain $c = 0.017$ which is consistent with the markedly asymmetric behavior of the sentiment observed over the studied period.

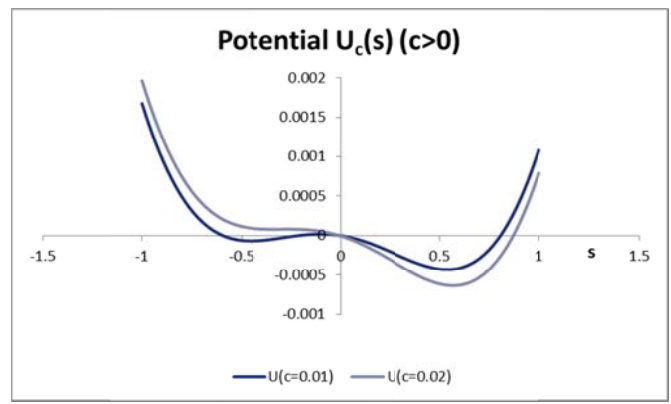

**Figure 4**: The potential $U_c(s)$ for $\beta_1 = 1.1$ and $c = 0.01, 0.02$. Only one equilibrium point, which corresponds to the minimum of $U_c(s)$ at $s = s_+$, survives the distortion of the potential well for $c = 0.02$.

---

[14] The potential $U_c(s)$ can be expanded into a truncated Taylor series and written as $U_c(s) = -w_s \left( \frac{\beta_1 - 1}{2} s^2 - \frac{\beta_1^3}{12} s^4 + cs \right)$. Note that the asymmetry relative to the origin ($s = 0$) emerges via the third term as it is linear in $s$.

Lastly, we would like to mention that despite similarities in the behavior between sentiment and the S&P 500 Index, the differences are nevertheless substantial. Perhaps most importantly, the evolution of sentiment is bounded whereas the stock market is, on average, experiencing growth. On the other hand, it seems reasonable to suppose that all market developments, including its long-term growth, occur through developments in sentiment. To find whether or not this is true, we must understand the relation between investor sentiment and market price, which is the subject of the next section.

## 1.3. Price

In this section we will be concerned with investigating how sentiment *s* can impact market price *P*. We propose a model of price formation and apply it to construct an empirical time series of model prices that we will compare with observed market prices.

### 1.3.1. Model of price formation

Positive or negative sentiment means that on average investors believe that the market will rise or fall, respectively. But does it also imply that investors would necessarily act on their sentiment and proceed to buy or sell assets, as the case may be?

Let us consider an investor who has just allocated capital to a stock market. Next day, all other things being equal, the investor is unlikely to increase or decrease her market exposure unless her sentiment has changed because she had already deployed capital in the amount reflecting that same level of sentiment. It is therefore sensible to assume that, ignoring external constraints (e.g. capital, risk, diversification), investment decisions are mainly driven by the change in sentiment on timescales where the investor's memory of past sentiment levels persists, determined by $\Delta t \ll \tau_s$ (see Section 1.2.2).

Reasoning similarly, we conclude that on longer timescales ($\Delta t \gg \tau_s$) investors would invest or divest based primarily on the level of sentiment itself because their previous allocation decisions would no longer be linked in their memory to the previous levels of sentiment (recall that our initial guess for $\tau_s$ is approximately 1 month).

As the change of market price is determined by the net flow of capital in or out of the market, we conclude that price changes would depend primarily on sentiment changes on timescales shorter than $\tau_s$ (i.e. probably days to weeks) and on sentiment itself on timescales longer than $\tau_s$ (i.e. probably months to years), that is

$$\dot{p} \sim \dot{s}, \text{ i.e. } p \sim s \qquad \text{for } \Delta t \ll \tau_s, \tag{8a}$$

$$\dot{p} \sim s, \text{ i.e. } p \sim \int s dt \qquad \text{for } \Delta t \gg \tau_s, \tag{8b}$$

where $p$ is the logarithm of price $P$, i.e. $p = \ln P$.[15]

We reiterate that equations (8a) and (8b) provide only the asymptotic relations between sentiment and price at respectively short and long timescales. As we do not know the form of the actual equation governing the evolution of price, we simply superpose both asymptotic views and seek price $p$ at any time $t$ as

$$\dot{p} = a_1 \dot{s} + a_2 s + a_3, \quad \text{or equivalently } p(t) = a_1 s + \int (a_2 s + a_3) dt + a_4, \tag{9}$$

---

[15] By taking the logarithm of $P$, we normalize the price so that $s$ or $ds$ results in the relative price change $dp = d(\ln P) = \frac{dP}{P}$, i.e. the return, which is a standard procedure. If the price had not been normalized, $s$ or $ds$ would have caused different percentage changes in the price depending at which price level it occurs.

in the hope that solutions given by it approximate true solutions reasonably well. Note that constants $a_1$ and $a_2$ are positive, whereas constants $a_3$ and $a_4$ can take any sign.

Equation (9) can be written as

$$\dot{p} = a_1 \dot{s} + a_2 (s - s_*), \tag{10}$$

where $s_* = -\frac{a_3}{a_2}$.

Equation (10) implies that the change in price at $\Delta t \gg \tau_s$ is proportional to the deviation of sentiment from a certain value given by $s_*$. Thus $s_*$ in (10) serves as a yardstick, averaged across the investment community, relative to which investors appraise sentiment: if sentiment is above or below it, they may invest or divest, respectively. A nonzero $s_*$ can be interpreted as an implied reference sentiment level that investors are accustomed to and consider normal.[16] We do not wish to impose any a priori constraints on $s_*$. Instead we will determine its value by fitting price observations to the model in the next section.

### 1.3.2. Modeled price

In this section we apply equation (10) to construct the model price $p(t)$ from the time series $s(t)$ reported in Figure 3.

We search for the coefficients in equation (10) that minimize the mean-square deviation between $p(t)$ and the log prices of the S&P 500 Index over the period covered by our data. Figure 5a, which depicts the evolution of model prices vs. index prices, shows that $p(t)$ approximated the index behavior over this period well. We note, first, that the daily model prices and the daily index

---

[16] A nonzero $s_*$ would be compatible with the adaptation level theory put forward by Helson (1964) and the reference point concept introduced by Kahnemann and Tversky (1979) in their prospect theory.

Page **27** of **97**

prices are 83% correlated and, second, that standard tests show that these variables are cointegrated. Additionally, Figure 5b depicts the behavior of the components of $p(t)$ (eq. 8). As expected, the leading component at $\Delta t \ll \tau_s$ (eq. 8a) exhibits rapid, small-to-mid-scale fluctuations while the leading component at $\Delta t \gg \tau_s$ (eq. 8b) exhibits slow, large-scale variation with time.

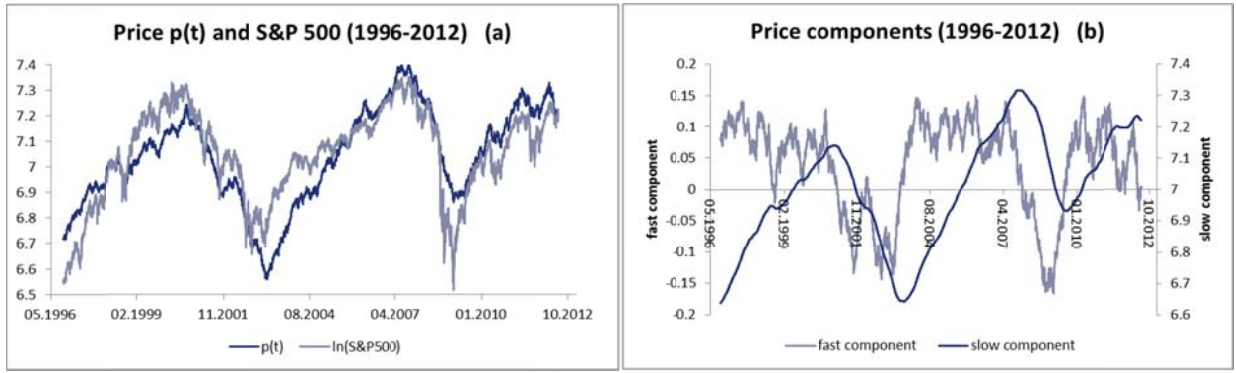

**Figure 5**: (a) The model price $p(t)$ vs. the S&P 500 log prices (1996-2012). The price $p(t)$ is obtained from eq. (10) with coefficients estimated using the least squares fitting and $s(t)$ from Fig. 3. The estimated coefficients are $a_1 = 0.322;. a_2 = 0.003; a_4 = 6.564;$ and $s_* = 0.076$. (b) The fast and slow components of $p(t)$ given by eq. (8a) and eq. (8b), respectively.

We note that the constant $s_*$ is positive over this period. Indeed it would be reasonable to expect that $s_*$ stays on average positive as it manifests a background reference value that investors consider normal and so is likely to be related to the positive, long-term bias exhibited by the sentiment (see Section 1.2.3).

It is shown in Appendix B that in the leading order $s_*$ coincides with the equilibrium value of sentiment (as is expected given its context as the background reference value). We know that the equilibrium value is determined solely by order parameter $\beta_1$, which is the inversely proportional to temperature $\theta$.[17] This means that temperature has a direct impact on price dynamics. As temperature indicates the balance between random behavior and herding behavior in the market, it would be sensible to presume that such balance may gradually shift over time and consider $\theta$ to be slowly varying with time. Consequently, we can expect the reference sentiment level $s_*$ to be a slowly varying function of time as well.

Assuming that $\theta = \theta(t)$ we calculate $p(t)$ from equations (4) and (10) using an iterative process for minimizing the mean-square deviation between the model and the index for $\beta_1 = \beta_1(\theta(t))$ and $s_* = s_*(\theta(t))$, while keeping other parameters constant.[18] As a result, we obtain a better fit between the model and the index: now $p(t)$ matches the index behavior at various timescales more closely (Fig. 6) and the correlation between the daily model prices and the daily index values has improved to 97%.

---

[17] The equilibrium points are determined by the condition that $H = 0$ and $\dot{s} = 0$ in equation (4) (or equivalently by the extrema of $U_0(s)$ (eq. 6)), given by the equation $s = \tanh(\beta_1 s)$, so that the equilibrium value depends only on $\beta_1(\theta)$.

[18] As follows from equation (10), the reference sentiment level $s_*$ is an important factor for price dynamics. In the leading order $s_*$ is exclusively a function of $\beta_1(\theta)$: $s_* = s_*(\beta_1(\theta))$ (Appendix B). As we are interested in the effect of a slowly varying $\theta(t)$ on $s_*$, we consider $\beta_1$ to be a function of time but for simplicity hold other parameters constant.

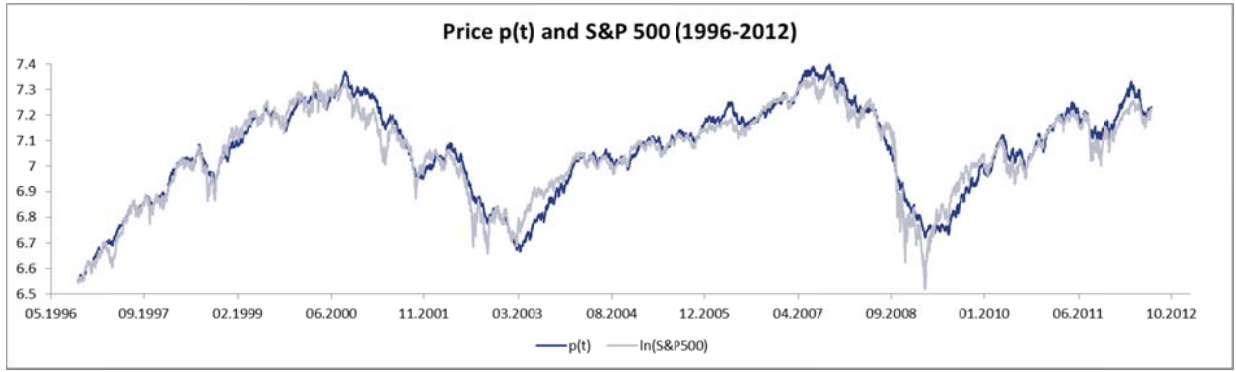

**Figure 6**: The model price $p(t)$ vs. the S&P 500 log prices (1996-2012). The price $p(t)$ is constructed through an iterative least squares fitting involving eq. (4) and (10), in which $\beta_1 = \beta_1(\theta(t))$ and $s_* = s_*(\theta(t))$ and the relation between $s_*$ and $\theta$ is determined from eq. (B5) as shown Appendix B. The estimated mean parameter values: $\beta_1 = 1.12$ and $s_* = 0.173$. The estimated values of the constant coefficients (eq. 10) are $a_1 = 0.376$; $a_2 = 0.003$; and $a_4 = 6.293$. The values of $w_s$ and $\beta_2$ are shown in Fig. 3.

Let us now consider $\theta = \theta(t)$ obtained using the above-outlined iterative process and reported in Figure 7. We observe that $\theta(t)$ exhibited small variations, always remaining lower than unity. These values of $\theta$ correspond to a slightly supercritical $\beta_1$ (i.e. $\beta_1 \gtrsim 1$), which indeed fluctuated around the average value of 1.1, maintaining a value above unity at all times. It follows that the sentiment evolved, as expected, inside the double-well potential and that herding behavior was prevalent throughout the studied period, although its influence tended to ebb and flow with time.

We observe that the periods of higher temperature correspond to bear market regimes (2001-2002 and 2008) and that the periods of lower temperature correspond to bull market regimes (2003-2007) with the notable exception of the post-2008-crisis market rally, captured in our data for 2009-2012, that took place in a relatively high temperature environment [19].

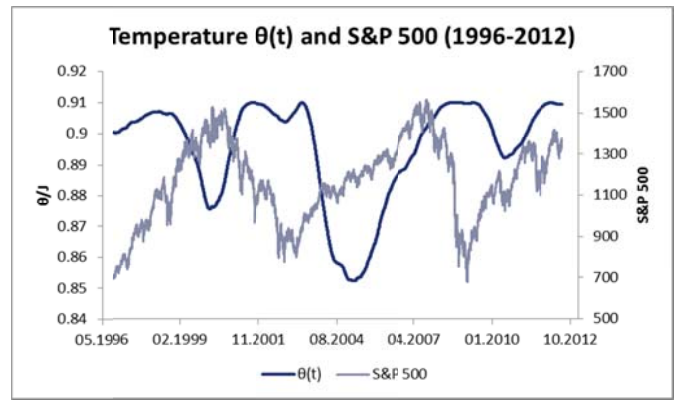

**Figure 7**: Temperature $\theta$ (measured in units of $J$ (eq. 3)) as a function of time (1996-2012). Temperature fluctuated less than 10% around the mean value of 0.9, always staying near but below the critical value of unity, which marks the phase transition between ordered and disordered states.

As periods of higher temperature indicate increased importance of random behavior, it would be logical to suppose that these periods coincide with periods of increased volatility in investor sentiment (expressed differently: rising temperature flattens the potential well, which makes it easier for information flow to move sentiment). Figure 8a demonstrates an evident correspondence between these two variables, while Figure 8b shows similarities in behavior between temperature and market volatility, which is also to be expected.

---

[19] Figure 7 also shows elevated temperatures during the market rally of the late 1990s. However, the temperature pattern prior to 2000 may be unreliable due to low data volumes. Furthermore, the subsequent decline in temperature in 1999-2000 suggests that the levels of 1996-1998 could be an artifact of low data volumes for that period.

Note that the post-2008-crisis bull market exhibited an unexpectedly high sentiment volatility in 2010-2012, which is compatible with the above observation that the temperature did not decrease following the Credit Crisis. A reasonable explanation to it is that, unlike the bull market in 2003-2007, the bull market during 2010-2012 was basically a liquidity-driven rally occurring in the conditions of subdued economic recovery and high uncertainty in overall investor sentiment (the risk-on, risk-off environment).

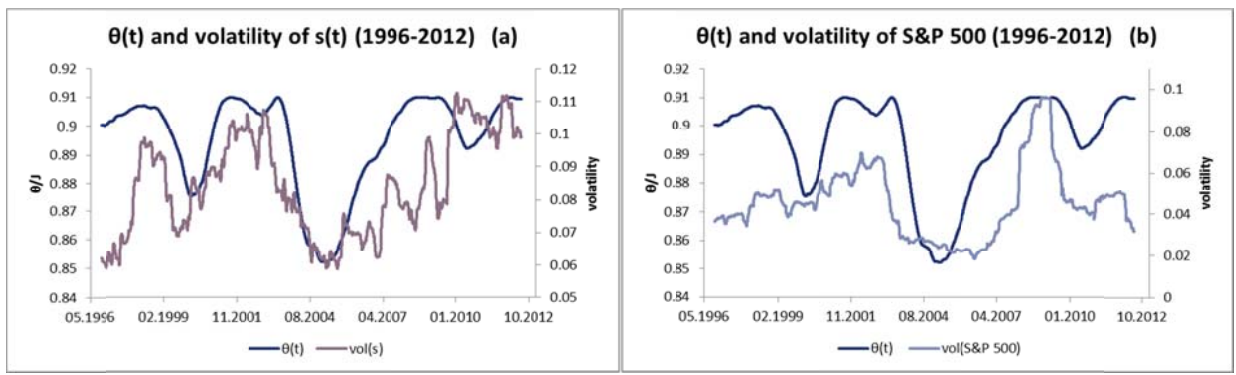

**Figure 8**: (a) Temperature $\theta$ and the volatility of sentiment $s$ (1996-2012). Sentiment volatility is calculated as the standard deviation of monthly (21 days) increments in $s$ over a rolling 300-day period anchored to the date for which the volatility is computed. (b) Temperature $\theta$ and S&P 500 Index volatility over that same period. S&P 500 Index volatility is calculated as the standard deviation of 21-day log returns using the same method as in (a).

In summary, we have demonstrated that the model can, within a relatively close tolerance, reproduce market price behavior as a function of investor sentiment. Additionally, we have shown that the balance between herding behavior and random behavior, indicated by the economic analog of temperature $\theta$, can gradually change over time. As we will see in Section 2, $\theta$ plays an important role in influencing market dynamics.[20]

For convenience, the estimates of parameters are reported in the table below.

**Table I**
**Parameters applied for modeling $s(t)$ and $p(t)$**

| Parameter | Value | Source |
|---|---|---|
| $\bar{H}$ | 0.017 | DJ/Factiva (Section 1.1) |
| $\sigma_H$ | 0.408 | DJ/Factiva (Section 1.1) |
| $\tau_s$ | 25 (business days) | Initial guess (Section 1.2.3) |
| $w_s = 1/\tau_s$ | 0.040 | |
| $\beta_1$ | 1.100 | Average $\beta_1$ (Fig. 7) |
| $\beta_2$ | 1.000 | Initial guess (Section 1.2.3) |
| $a_1$ | 0.374 | Least squares fitting Eq. 10 |
| $a_2$ | 0.002 | Least squares fitting Eq. 10 |
| $a_4$ | 6.500 | Least squares fitting Eq. 10 |
| $s_*$ | 0.131 | Least squares fitting Eq. 10 |
| $c = \beta_2 \bar{H}$ | 0.017 | |

---

[20] A comment relevant to the following sections: We apply non-constant $\beta_1 = \beta_1(\theta(t))$ and $s_* = s_*(\theta(t))$ for two purposes, to analyze the variation of $\theta$ with time and to demonstrate that by accounting for this variation we achieve a closer fit between the model and the market. Since the variation of $\theta$ is small and for the sake of simplicity, we will henceforth use the time series $s(t)$ and $p(t)$ as calculated with constant parameters. The values of parameters are shown in Figures 3 and 5, except that we apply $\beta_1 = 1.1$ (the average value of $\beta_1(t)$ over the studied period) instead of $\beta_1 = 1.0$. The parameter values are also summarized in Table I, however the graphs $s(t)$ and $p(t)$ are not shown for the economy of space and due to their similarity to those depicted in Figures 3 and 5, respectively.

## 1.4. Discussion

In this section we summarize the findings of the empirical study and discuss their theoretical implications and practical applications.

### 1.4.1. Results

We have shown that the model price replicated the S&P 500 Index values over the period 1996-2012 within reasonable tolerance, lending credence to the assumptions made early on to develop the model. Thus we conclude:

First, investors' sentiment is influenced by direct information flow.

Second, in addition to direct information, sentiment is also influenced by the interaction among investors and idiosyncratic influences that can be assumed random for our purposes. The Ising model, in which temperature serves as a measure of random influences, provides the relevant framework for studying sentiment dynamics and, in particular, enables us to obtain a closed-form equation for the evolution of sentiment.

Third, the mechanism of price formation as a function of sentiment works differently on different timescales. On timescales shorter than the investors' average memory horizon, market price changes proportionally to the change in sentiment. On longer timescales, price changes proportionally to the deviation of sentiment from a reference level that is generally nonzero. As a result, price development is naturally decomposed into a slow, large-scale variation with time and fast, small-to-mid-scale fluctuations.

We have seen that for the studied period the effects in connection with herding behavior of investors were more prevalent than those related to random behavior. Consequently, sentiment evolved inside a W-shaped potential, with a negative equilibrium value in one well and a positive value in the other. Direct information flow, which acts as a force moving sentiment inside these wells, was on average positive during the studied period. Two results became evident: first, sentiment spent most of the time in the positive well and crossed into the negative well only during the time of crises and, second, the reference sentiment level stayed mostly positive during this period.

Although our data coverage is limited to 16 years, we tend to think that the above-described asymmetry of sentiment dynamics, caused by the positive bias in direct information flow, may generally persist. This is because direct information must be on average positive to be consistent with the long-term growth of the stock market. The asymmetric behavior of sentiment can be alternatively explained in terms of the perturbation of potential: the constant proportional to the (positive) mean of direct information flow is a parameter that distorts the shape of potential, making the positive well deeper than the negative well. As a result, sentiment has a propensity to be on average positive as it takes more significant news to turn it from positive to negative than from negative to positive.

We would like now to comment on certain effects pertaining to the long-term (months to years) behavior of sentiment and price. We begin with the economic analog of temperature, which determines the relative importance of random behavior vs. herding behavior. Temperature can substantially impact market dynamics in two ways: first, by altering the shape of potential thereby affecting sentiment evolution (e.g. the probability of crossing the wells) and, second, by influencing the reference sentiment level (eq. 10) which is sensitive to temperature changes.

We have observed that temperature and the volatility of sentiment varied over time in a similar pattern. It seems reasonable to assume that periods of increased sentiment volatility imply a high uncertainty of investor opinion symptomatic of bear markets and risk-on, risk-off regimes[21], while periods of low sentiment volatility correspond to the positive, trending markets. We note that the liquidity-driven market recovery during 2010-2012 appears to be an exception to this observation, as it exhibits typical risk-on risk-off features, exemplified by the elevated sentiment volatility (Fig. 8a).

We now briefly consider a connection with the economy. Figure 9a,b shows that temperature varied in a pattern resembling economic fluctuations, typically increasing during the economic downturn and decreasing during the upturn. Figures 9c,d show that direct information flow and sentiment also fluctuated approximately synchronously with economy. These observations imply that the model captures some effects of the interaction between the general economy and the stock market.

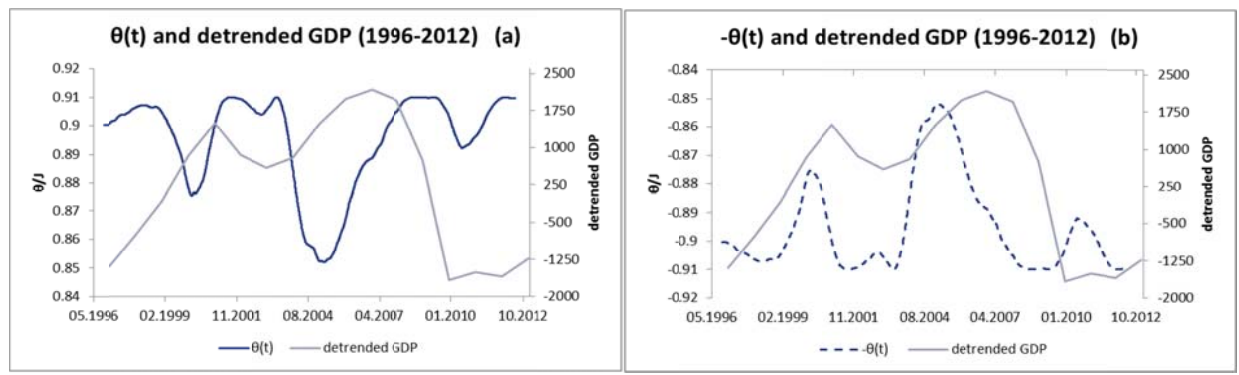

---

[21] "Risk-on, risk-off" is a loosely defined term that practitioners use to describe markets characterized by high volatility of the risk appetite of investors. It is a widely held view across the investment industry that during 2010-2012 the market was in a risk-off risk-on regime.

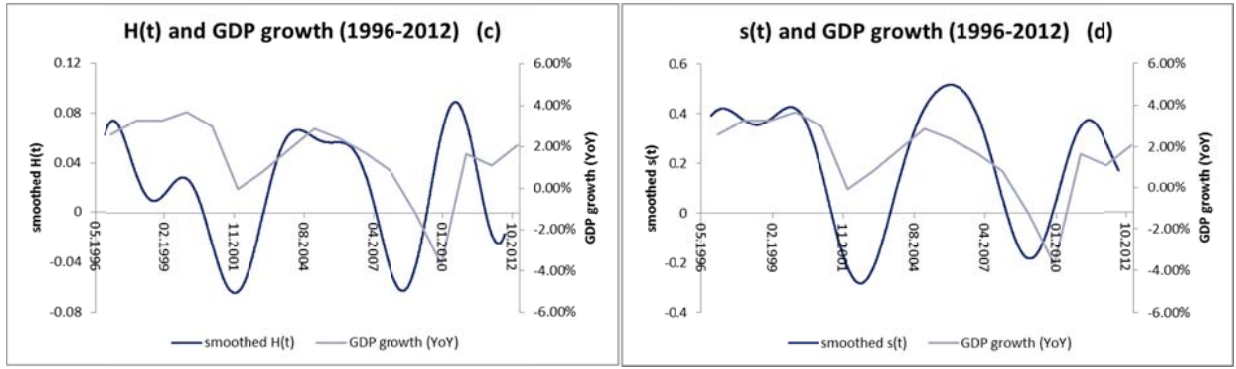

**Figure 9**: (a) Temperature $\theta(t)$ and the detrended US real GDP per capita (1996-2012). (b) The mirror image of $\theta(t)$ and the same GDP graph as shown in (a) for ease of comparison. (c) Direct information flow $H(t)$, smoothed using a Fourier filter that removed harmonics with periods less than 850 days, and the US real GDP per capita growth (YoY) (1996-2012). (d) Sentiment $s(t)$, similarly smoothed as $H(t)$ above, and the same GDP graph as shown in (c).

Finally, we have not yet touched upon the issue of whether or not this model is predictive. This will be done, briefly, in the next section, in which we highlight two relevant properties of the model. We will revisit this issue in the second part of this paper (Section 2), which further develops the theory underlying this model, emphasizing the potential for market forecasting.

### 1.4.2. Practical applications

We have observed two types of behavior characteristic to sentiment: the mid-term bounded motion inside the positive well and the infrequent crossings between the wells. The former seems to induce the mid-term market trends, while the latter appears to drive the large-scale dynamics of market declines and recoveries. It would be interesting to investigate whether there is some correspondence between the turning points of sentiment trends and market price trends.

A consideration of equation (10) leads to a conclusion that the turning points of sentiment evolution ($\dot{s} = 0$) and the turning points of price trends ($\dot{p} = 0$) never coincide unless the second term in (10) is zero ($s = s_*$). It follows that in certain cases the turning points of sentiment trends can occur before the turning points of price trends. For example, in situations where an upward trend in sentiment is reversed at $s > s_*$, $\dot{s}$ changes sign from positive to negative, while $\dot{p}$ continues to be positive, as coefficients $a_1$ and $a_2$ in equation (10) are positive. This means that sentiment reverses the trend in advance of the market's reversal. Similarly, when a downward trend is reversed at $s < s_*$, $\dot{s}$ changes sign from negative to positive, while $\dot{p}$ remains negative, such that sentiment again breaks the trend before the market.

To verify the above conclusion empirically, we extract two first quasiperiodic components of the multivariate singular spectrum analysis (MSSA) expansion of the sentiment $s(t)$ and the S&P 500 Index and plot them in Figure 10a. The index can be seen to trail sentiment with a considerable lag. Figure 10b shows the same MSSA expansion for the slow component of price $p(t)$, given by the second term in equation (10), and the S&P 500 Index. It is evident that the slow price component trails the index. This compensates for the lead that sentiment (i.e. the first term in equation (10)) has over the index, and, as a result, the overall model price is not substantially shifted relative to the index on long timescales. These observations may be helpful for the development of long-term momentum strategies.

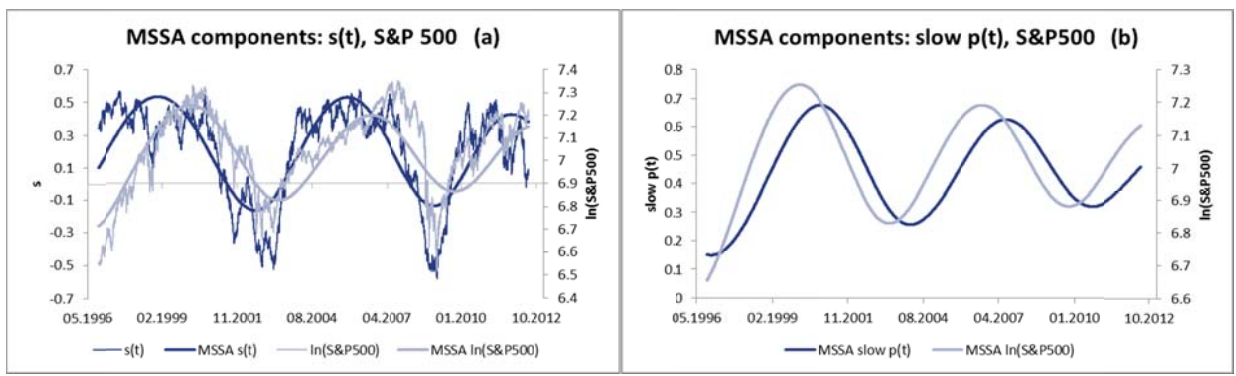

**Figure 10**: MSSA expansion (1996-2012) (a) the sentiment $s(t)$, which is proportional to the fast price component (eq. 8a), and the S&P 500 log price; and (b) the slow price component (eq. 8b) and the S&P 500 log price.

Next we consider the possibility of predicting the evolution of sentiment itself. In Section 1.2.3 we have pointed out that such possibility exists in principle, provided that direct information flow $H(t)$ is not white noise, i.e. its autocorrelation is not a delta function. Figure 11 shows that the autocorrelation of the daily measured $H(t)$ does not behave as a delta function and is noticeably different from the autocorrelation of the S&P 500 log returns over that same period (the pattern of which indeed makes a compelling argument for the random walk hypothesis). Thus, we conclude that it could in principle be possible to predict the behavior of $s$ and therefore $p$ from the time series $H(t)$. Further research in this area may prove useful to market practitioners. We will return to this discussion in the second part of this paper (Section 2).

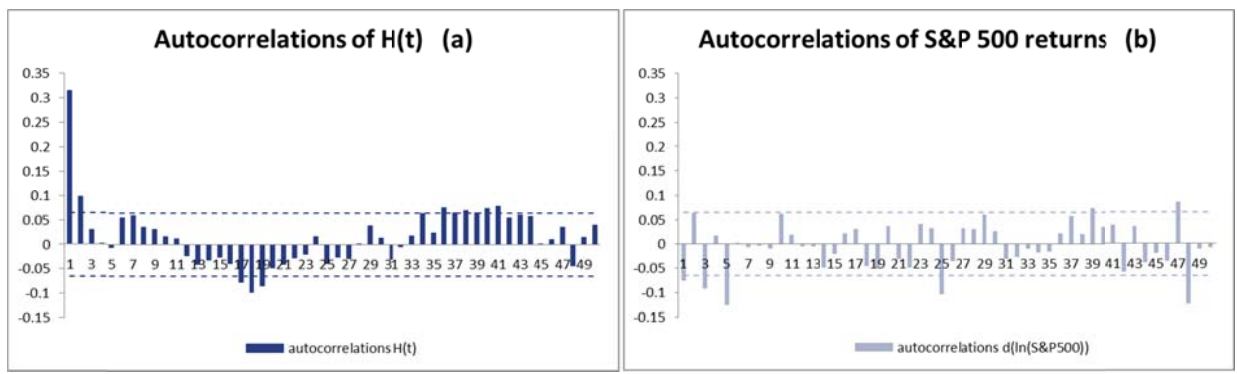

**Figure 11**: (a) Autocorrelations of daily $H(t)$ across different time lags. We take $H(t)$ obtained from the proprietary news archive covering the period 2010-2012 as it has a clearer signal due to higher daily data volumes than $H(t)$ based on the DJ/Factiva archive (see Fig. 1b). (b) Autocorrelations of daily log returns of the S&P 500 Index across different time lags for the same period. The confidence intervals are also indicated in the graphs.

It is worthwhile noting that the autocorrelation pattern of $H(t)$ resembles the autocorrelation of a process that contains decaying quasiperiodic oscillations, with the caveat that the autocorrelation pattern at time lags greater than 2 days is mostly within the margin of error. This observation is consistent with the results produced by the extended, self-contained model we develop in the next section.

## 2. Part II – Theoretical study of stock market dynamics

In the empirical study we have assumed that information that speculates about future market returns (direct information) may be effective in influencing investors' opinions about the market (investors' sentiment). We have established a mechanism that translates direct information via investors' sentiment into market price. The translation mechanism itself is mathematically simple in the sense that the equation for sentiment dynamics (eq. 4), which defines how investors form opinions based on received information, and the equation for price formation (eq. 10), which defines how investors make investment decisions based on their opinions, constitute an uncoupled and integrable system of equations.

It is therefore not surprising that the source of complex behavior, characteristic to actual markets, in the model is contained within the direct information flow that we have treated as exogenous. This means that we have investigated just one part of the problem and to learn more about the origins of market behavior we have to extend the framework by incorporating into it assumptions on how direct information flow is generated and channeled in the market.

In this part of the study we develop a theoretical (self-contained) model of the stock market that includes direct information $H$ (which we will call $h$ to set it apart from the empirical case) as a variable, alongside sentiment $s$ and price $p$. This model is simplified by construction to facilitate its study, so its solutions cannot reproduce the full range and the precise detail of actual market regimes. Its purpose is to capture the essential elements of market behavior and explain them in terms of the interaction among direct information, investor sentiment and market price.

This part proceeds as follows: Section 2.1 develops a self-contained model of stock market dynamics and Section 2.2 provides the analysis, results and discussion. The relevant technical details are in Appendices A and C.

## 2.1. News-driven market model

Direct information, which is based on the interpretation of general news, consists of opinions about future market performance that can reach a large number of investors in a short period of time. Such opinions can come from financial analysts, economists, investment professionals, market commentators, business news columnists, financial bloggers – basically, anyone with a view about the market, the means to deliver it to a large investor audience and the credibility to instill trust. For convenience, let us refer to them collectively as market ***analysts***. In the context of our framework, analysts perform a vital function: they interpret news from various sources, opine how the market might react and transmit their views through mass media. That is, they convert any type of information into direct information and make it available to market participants.[22]

---

[22] We consider only publicly available direct information simply because we know how to measure it. Usually, prior to an expert opinion being expressed publicly, it is discussed in professional circles and diffuses across the investment community. Thus, it would be normal that privately expressed views would have a lead over those which are publicly expressed. We tend to think, however, that on the timescale of investment decision-making by institutional investors, this informational lag is negligible due to the speed with which information becomes public through online publication, email and blogs. The topic of how opinions transit from private to public is deserving of its own research, but is outside the scope of this work. Here we assume that, for our purposes, the time lag between the public and private direct information can be neglected while their magnitudes (as measured by $H$) can be considered proportional.

Let us construct an Ising-type model with two types of interacting agents: investors who have capital to invest but cannot interpret news and analysts who can interpret news but do not have capital to invest (the irony of which is not lost on us). At any point in time, each investor and each analyst has either a positive (+1) or negative (-1) opinion about future market performance – the binary market sentiment $s_i$ in the case of investors and the binary market sentiment $h_j$ in the case of analysts. Investors can interact with each other and with analysts and, similarly, analysts can interact with each other as well as with investors. They interact by literally imposing opinions on each other as every sentiment tries to align other sentiments along its directional view. It is important to note that although the nature of interaction is the same (the exchange of opinions), the means of interaction in the model are very different: analysts, who are outnumbered by investors, are assumed to exert a disproportionally strong force on investors due to access to mass media.

We can now replace the homogeneous, single-component Ising system (1), where the external flow of direct information acted on investors, by its heterogeneous, two-component extension, in which investors and analysts (instead of information flow) interact with each other. We define $\{s_i\}$ and $\{h_j\}$ as individual sentiments of investors and analysts, respectively, and $N_s, N_h$ as their numbers ($N_s \gg N_h \gg 1$) and, as before, write the system's total energy (Appendix A, eq. (A1)) and apply the all-to-all interaction pattern to obtain in the limit $N_s \to \infty, N_h \to \infty$ the dynamic equations that describe the evolution of the variables as statistical averages, namely: $s = \langle s_T \rangle / N_s$ where $s_T = \sum_{i=1}^{N_s} s_i$ and $h = \langle h_T \rangle / N_h$ where $h_T = \sum_{j=1}^{N_h} h_j$.

The general form of the dynamic equations is as follows (Appendix A, eq. A14):

$$\dot{s} = -w_s s + w_s \tanh\left(\frac{J_{11}s + J_{12}h + \mu_s b_s(t)}{\theta}\right), \tag{11a}$$

$$\dot{h} = -w_h h + w_h \tanh\left(\frac{J_{21}s + J_{22}h + \mu_h b_h(t)}{\theta}\right), \tag{11b}$$

where, analogous to equation (3), the coefficient $J_{11}$ defines the strength of interaction among the investors; $J_{12}$ – the strength with which the analysts act on the investors; $J_{21}$ – the strength with which the investors act on the analysts; $J_{22}$ – the strength of interaction among the analysts; $b_s(t)$ and $b_h(t)$ are any general external forces, so far undefined, acting respectively on the investors and the analysts, where $\mu_s$ and $\mu_h$ determine their impacts; $w_s = 1/\tau_s$ with $\tau_s$ being the characteristic horizon of the investors' memory and $w_h = 1/\tau_h$ with $\tau_h$ being the characteristic horizon of the analysts' memory; and $\theta$ has been identified in equation (3) to be the economic analog of temperature or the parameter that determines the impact of random influences and consequently the level of disorder in the system. Note that parameters $J_{11}, J_{12}, J_{21}, J_{22}, \mu_s, \mu_h, w_s, w_h$ and $\theta$ are non-negative constants.

In our framework, analysts must react to new information faster than investors, so that their resistance to a change in sentiment is weaker than that of investors. Accordingly, we assume $\tau_h$ to be an order of magnitude smaller than $\tau_s$, resulting in $\tau_h \approx 2.5$ business days, This value is consistent with the behavior of the autocorrelation of $H(t)$ (Fig. 11a) that shows a rapid decay of "memory" effects on the order of 1-3 business days. Hence we set $w_h = 0.4$.

Let us consider the external forces $b_s(t)$ and $b_h(t)$, which have the meaning of external sources of information tapped by investors and analysts, respectively. We set $b_s = 0$ by assuming that investors in the model receive direct information only through analysts; in other words, investors' sentiment $s$ may change only by interacting with analysts' sentiment $h$.

The case of $b_h$ is more interesting because of the analysts' function to translate news into direct information and so provide a channel through which exogenous information enters the model. Since any news may be of relevance in this context – e.g. economic data release, corporate news, central bank announcements, political events or, say, change of weather conditions – we can make the usual assumption that new information arrives randomly and therefore represent news flow via noise.

There is, however, a particular source of information that should be considered separately. It is the change of market price itself.[23] Its importance is supported by empirical evidence: today's values of $H(t)$ are correlated with yesterday's and today's S&P 500 log returns at over 30% and 50%, respectively.[24]

Thus, we divide information available to analysts in the model into information related to the change in market price and all other (external) information. We write $b_h = \rho_1 \dot{p} + \rho_2 \xi(t)$, where $\dot{p}$ is market price change (eq. 10), $\xi(t)$ is random news flow (noise) and $\rho_1$ ($\rho_1 \geq 0$) along with $\rho_2$ are scaling constants. Note that by writing $\dot{p}$ in the expression for $b_h$, we have assumed that the feedback in the form of price change is continuous and contemporaneous, which is consistent with the fact that prices can be observed – directly as data – at any time and in real-time.

---

[23] Feedback in connection with price observations has been implemented in several agent-based models. For example, Caldarelli, Marsili and Zhang (1997) used the rate of change in price along with its higher-order derivatives, Bouchaud and Cont (1998) applied price deviations from fundamental values, and Lux and Marchesi (1999, 2000) considered both the rate of change in price and price deviations from fundamental values. The idea that price observations create a feedback loop which can produce complicated dynamics, leading to market rallies and crashes, has roots in the 19th century (see a review by Shiller (2003)). With regard to early feedback models, we take particular note of Shiller's (1990) model with lagged, cumulative feedback operating over long time intervals, implying that information related to past price changes has long-lasting effects.

[24] Interestingly, the cross-correlation function is approximately zero at positive time lags. We could plausibly explain this by stating that on short timescales (e.g. intraday) there is an efficient market response to news, whereby new information is almost immediately reflected in prices as obvious price anomalies are swiftly arbitraged away by investors (e.g. the high frequency traders). The response to news on longer timescales has a different nature, and its mechanics are the subject of the present work.

We can now substitute the external forces $b_s = 0$ and $b_h = \rho_1 \dot{p} + \rho_2 \xi(t)$ into equations (11) and write the model in terms of $p, s$ and $h$ as follows:

$$\dot{p} = a_1 \dot{s} + a_2(s - s_*), \tag{12a}$$

$$\dot{s} = -w_s s + w_s \tanh(\beta_1 s + \beta_2 h), \tag{12b}$$

$$\dot{h} = -w_h h + w_h \tanh(\beta_3 s + \beta_4 h + \kappa_1 \dot{p} + \kappa_2 \xi(t)), \tag{12c}$$

where for convenience we have regrouped the constants to have $\beta_1 = \frac{J_{11}}{\vartheta}$, $\beta_2 = \frac{J_{12}}{\vartheta}$, $\beta_3 = \frac{J_{21}}{\vartheta}$, $\beta_4 = \frac{J_{22}}{\vartheta}$, $\kappa_1 = \frac{\mu_h \rho_1}{\vartheta}$ and $\kappa_2 = \frac{\mu_h \rho_2}{\vartheta}$.

We wish to make the initial investigation easier by further simplifying equation (12c). First, let us assume that the influence of investors on analysts ($\sim \beta_3$) and the interactions among the analysts ($\sim \beta_4$) are less important in respect of opinion making than the impact due to market performance ($\sim \kappa_1$) and exogenous news flow ($\sim \kappa_2$), and therefore we neglect the corresponding terms in (12c). Second, let us approximate $\dot{p}$ in (12c) as $\dot{p} = a_1 \dot{s} + r$, where $r$ is a positive constant that carries the meaning of the long-term growth rate of the stock market. That is, we replace $a_2(s - s_*)$ by $r$ when we substitute $\dot{p}$ from (12a) into (12c). As we have seen in Section 1.3.2, $a_2(s - s_*)$ changes slowly with time as compared with $a_1 \dot{s}$, therefore the approximation is valid for sufficiently short time intervals. In other words, by making this approximation we neglect the long-term feedback effects on $h$ relative to the analogous short-term effects.

As a result, system (12) can be simplified as

$$\dot{p} = a_1 \dot{s} + a_2(s - s_*), \tag{13a}$$

$$\dot{s} = -w_s s + w_s \tanh(\beta_1 s + \beta_2 h), \tag{13b}$$

$$\dot{h} = -w_h h + w_h \tanh(\gamma \dot{s} + \delta + \kappa \xi(t)), \tag{13c}$$

where $\gamma = \kappa_1 a_1$ ($\gamma \geq 0$), $\delta = \kappa_1 r$ ($\delta \geq 0$) and $\kappa$ is simply $\kappa_2$ renamed.

The stock market model is now fully defined. System (13) represents the stock market as a nonlinear dynamical system driven by the random flow of exogenous news $\xi(t)$ (Fig. 12a). Note that for a given $h(t)$ equation 13(b) becomes equation (4) where $H$ is replaced by $h$. Thus, analyst sentiment is, as expected, synonymous with direct information flow and hence the empirical market model (eq. 4 and 10), studied in Section 1, is a particular case of the theoretical market model (eq. 13), as shown in Figure (12). In the next section we will see that model (13), although extremely simple, possesses some surprisingly interesting dynamics consistent with the behaviors observed in the empirical study as well as with the actual behavior of the stock market.

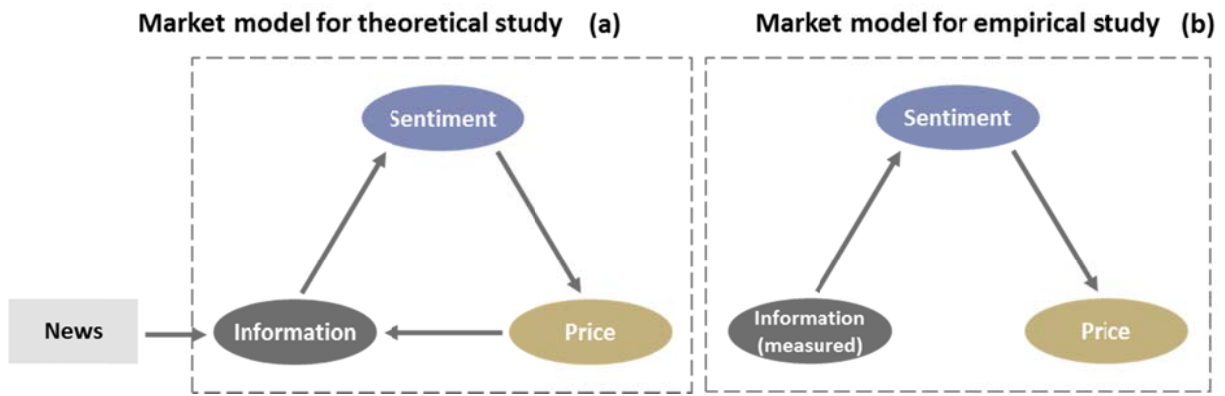

**Figure 12**: (a) A schematic view of the (simplified) theoretical model (eq. 13). Price $p(t)$ evolves according to (13a), describing how investors make investment decisions based on sentiment $s(t)$. Sentiment $s(t)$ evolves according to (13b) through interactions among investors and between investors and analysts. Direct information $h(t)$ evolves according to (13c) driven by the exogenous news flow $\xi(t)$ and by market performance approximated by the first two terms in the argument of the hyperbolic tangent in (13c). (b) A schematic view of the empirical model studied in Section 1. Direct information $H(t)$ is measured daily in the general news flow and then substituted into eq. (4) to obtain $s(t)$, which is subsequently substituted into eq. (10) to yield $p(t)$.

## 2.2. Model analysis and results

In this section we report and discuss the results of the study of system (13). We note that equations (13b,c) constitute a closed system that we can solve for $h(t)$ and $s(t)$ and subsequently substitute $s(t)$ into equation (13a) to obtain $p(t)$. We first consider the autonomous case $\xi(t) = 0$ and then study the time-dependent situation $\xi(t) \neq 0$.

### 2.2.1. Autonomous case: $\xi(t) = 0$

The two-dimensional nonlinear dynamical system (13b,c) is studied in detail in Appendix C. Here we report only the main results for the relevant range of parameter values.

At $\beta_1 = 1$ system (13b,c) experiences phase transition between the disordered state with one equilibrium point ($\beta_1 < 1$) and the ordered state with three equilibrium points ($\beta_1 > 1$). As before, we are interested in the ordered state in the vicinity of the phase transition ($\beta_1 \gtrsim 1$).

If $\delta = 0$, there is one unstable equilibrium point at the origin and two stable equilibrium points at $s = s_\pm$ located symmetrically with respect to the origin on the axis $h = 0$. If $\delta > 0$, the equilibrium points shift to new, asymmetric positions along the axis $h \approx \delta$ for $\delta \ll 1$. Once $\delta$ exceeds a critical value the distortion of the initially symmetric configuration becomes so strong that only the positive equilibrium point at $s = s_+$ survives.

In this respect the behavior of system (13b,c) is similar to that of the one-component sentiment model (4). To understand where a new behavior may originate, let us take a closer look at equations (13b,c). Direct information flow, described by analysts' sentiment $h$, causes investors' sentiment $s$ to change (eq. 13b). In turn, any change in $s$ also forces a change in $h$ as due to the feedback term $\gamma \dot{s}$ (eq. 13c). As a result, investors' sentiment is now coupled to direct information and vice versa, and it is this coupling that leads to the emergence of inertia and new behaviors related to it.

Indeed, as is shown in Appendix C, system (13b,c) can be approximated by

$$\ddot{s} + G(s,\dot{s})\dot{s} + \frac{dU(s)}{ds} = 0, \tag{14}$$

where the damping coefficient $G(s,\dot{s})$ and the potential $U(s)$ are given by equations (C15) and (C16), respectively. Equation (14) can be interpreted as the equation of motion for a damped particle of unit mass inside an asymmetrically shaped potential characterized by a deep well in which $s$ is positive and a shallow well in which $s$ is negative (Fig. 13d). The fact that the system has acquired a mass, and therefore inertia, results in a very different dynamic as compared to the one-component, inertialess model (4).

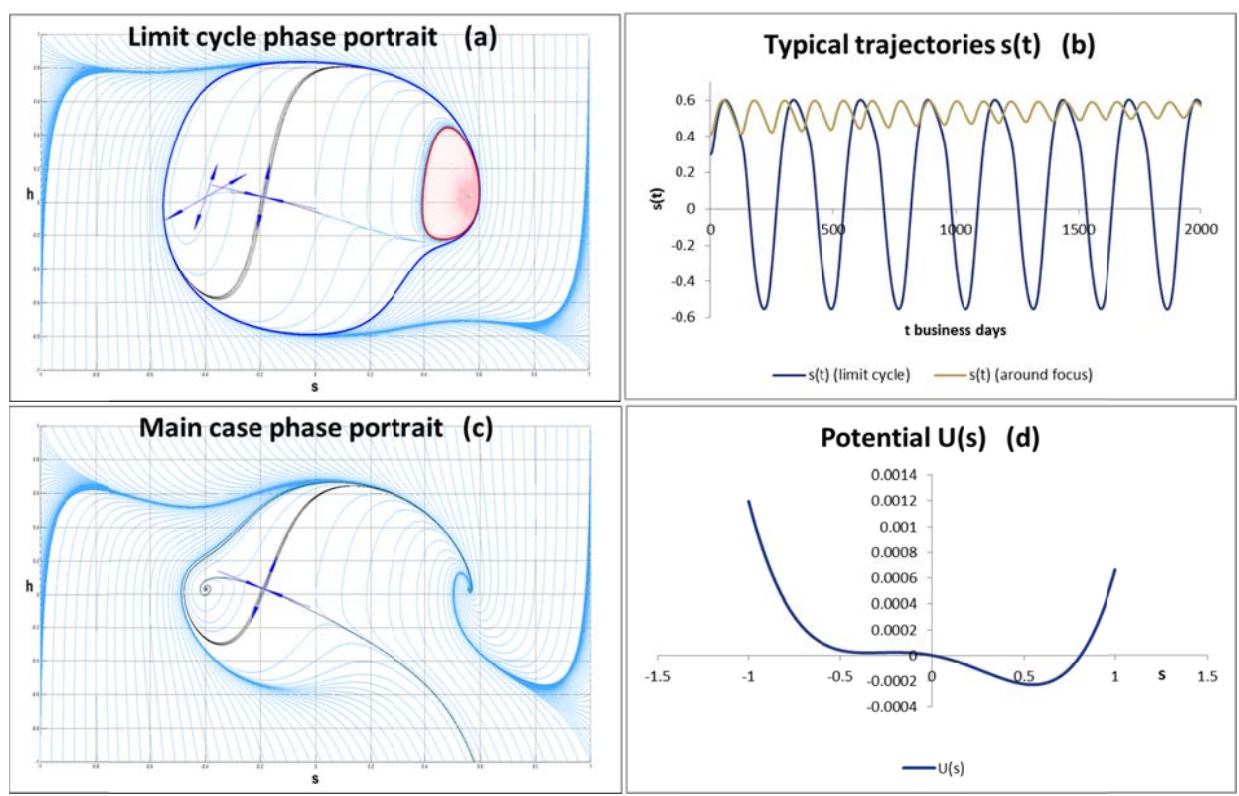

**Figure 13**: (a) The phase portrait ($\gamma = 67.7$) showing a large stable limit cycle (dark blue), a small unstable limit cycle (dark red) around a stable focus (green asterisk) in the deep, positive well ($s > 0$) and the equilibrium point in the shallow, negative well ($s < 0$) which is an unstable node. (b) The evolution of sentiment $s(t)$ in Fig. 13(a) along two typical trajectories, on the limit cycle and around the stable focus. (c) The phase portrait of the main case ($\gamma = 56.0$) showing an unstable

focus in the negative well (red asterisk), a stable focus in the positive well (green asterisk) and the trajectories passing near the saddle point located between the wells. (d) The potential $U(s)$ corresponding to (a) and (c). Parameters corresponding to Fig. 13(a-d): $\beta_1 = 1.1$; $\beta_2 = 0.55$; $\delta = 0.03$; $w_s = 0.04$; $w_h = 0.4$.

Most importantly, inertia enables free oscillations of $s$ and $h$, provided that the feedback coefficient $\gamma$ is not too small.[25] Figures 13a,b show two types of free oscillations that emerge in the system. The first type reflects decaying oscillations with a characteristic period of roughly 6 months around the equilibrium point inside the deep well. The second type reflects a large-scale stable limit cycle that carries sentiment from the deep well into the shallow well and back in approximately 14 months. Such large-scale motion is self-sustaining, as it is fuelled by the coupling of direct information and sentiment, such that large swings in $h$ cause large swings in $s$ that cause large swings in $h$ that cause large swings in $s$ and so forth.

Imagine that the flow of exogenous news were to be interrupted. Then, according to the model, two scenarios are possible. First, the market may converge to a steady state in which sentiment likely settles at the positive equilibrium value $s_+$ inside the deep well[26], while simultaneously the level of direct information approaches $\delta$. Alternatively, the market may go into a steady state in which sentiment and direct information are attracted to the limit cycle where they exhibit large-scale self-perpetuating oscillations between extreme negative and extreme positive values.

---

[25] When γ is increased from zero, it induces the following bifurcations of the equilibrium points at the bottom of each potential well ($s = s_\pm$): stable node -> stable focus -> unstable focus -> unstable node. Free oscillations become possible starting from the first bifurcation in the above sequence. Note that the equilibrium point at the cusp of the potential always remains an unstable saddle.

[26] It would be rather unusual if s were to come to rest at s₋ because, first, sentiment does not often cross into the shallow well and, second, such equilibrium state may be unstable or not exist at all, as the critical values of δ at which s₋ vanishes are within the admissible range of parameter values.

Figure 13c shows a regime immediately preceding the formation of the limit cycle. This regime is a prime candidate for generating a realistic sentiment behavior. It includes mid-term oscillations around the focus in the deep well, consistent with the bounded motion that we have observed in the empirical model and connected with mid-term market trends in Section 1.2.3. It also contains the large-scale trajectories, located in the vicinity of the (about-to-be-formed) limit cycle, that lead sentiment via quasi-self-sustaining motion into and out of the negative sentiment territory. The evolution of sentiment along these trajectories is consistent with its observed behavior at the time of market crashes and rallies (Section 1.2.3).

Lastly, we note that this regime, which we expect to be relevant to the real-world stock markets, exists under the realistic choice of parameter values consistent with the values obtained using the empirical data.[27]

---

[27] The parameters in the theoretical model (eq. 13 and Fig. 13c) have been chosen to coincide with the parameters of the empirical model (Table I), except that $\beta_2 = 0.55$ (theoretical) whereas $\beta_2 = 1.0$ (empirical). Additionally: 1. $\gamma = 56$ therefore $\gamma w_s = 2.2$, so that the feedback term $\gamma \dot{s}$ participates in the leading order dynamics in equation (13c) while not dominating other terms. 2. $\delta = 0.03$ so that the constant $c = \beta_2 \delta = 0.017$, which determines the distortion of the potential, is close to the value in the empirical model (Table I). 3. It follows from the definitions of $\gamma$ and $\delta$ in equation (13) that $\delta = \gamma r/a_1$. Using $a_1$ and $w_s$ from Table I and the daily average logarithmic growth rate $r = 2.24 \times 10^{-4}$ of the DJ Industrial Average Index since 1914 (interpolated using annual data), we estimate $\gamma r/a_1 \sim 0.034$ which is close to $\delta = 0.030$ in the theoretical model.

## 2.2.2. Non-autonomous case: $\xi(t) \neq 0$

The news flow $\xi(t)$ acts as a random force in system (13). To study its influence, we make the assumption that $\xi(t)$ is normally-distributed white noise with zero mean and unit variance.[28]

The system's behavior will depend significantly on the relative magnitude of the terms $\gamma \dot{s}$ and $\kappa \xi(t)$ in equation (13c). Indeed, if $\gamma|\dot{s}| \ll \kappa|\xi(t)|$, the dominant noise will drive $h$ and $s$ randomly around the potential well, leading to nearly stochastic behavior in price $p$. Conversely, if $\gamma|\dot{s}| \gg \kappa|\xi(t)|$, the system's dynamics will primarily consist of a piecewise motion along the segments of phase trajectories in the autonomous case, as on average $h$ and $s$ will travel far along a trajectory by the time the noise will have been able to displace them.[29] In this case, the resulting behavior of market price will have a discernible deterministic component to it.

We set $\kappa = 1$ as this permits both terms to participate in the leading order dynamics.[30] However, as $\gamma|\dot{s}|$ is a function of $h$ and $s$ (eq. 13b), the relative importance of these terms actually depends on the location of any given trajectory, which means that the above-discussed characteristic dynamics

---

[28] Note that in simulations the noise $\xi(t)$ is, strictly speaking, white only on a daily (weekly, monthly, etc.) basis, i.e. when $t$ can be expressed in multiples of business day. This is because $\xi(t)$ can have nonzero intraday autocorrelations due to intraday interpolation in the numerical scheme.

[29] This statement is only valid if the system is structurally stable (i.e. non-chaotic) when perturbed by noise, which is likely the case as we have not been able to find the traces of chaos (e.g. positive Lyapunov exponents) in it for nonzero $\xi(t)$ in the relevant range of parameter values. (As a side comment, we observed positive Lyapunov exponents in situations where system (13) was forced by a periodic function of time, instead of noise).

[30] $\kappa = 1$, so $\kappa|\xi(t)| \sim \kappa \sigma_\xi = 1$, where $\sigma_\xi$ is the unit standard deviation of $\xi$, therefore $\kappa|\xi(t)|$ has the same order of magnitude as $\gamma|\dot{s}| \sim \gamma w_s \lesssim 2.2$ for $|\dot{s}| \lesssim 1$.

may be bound to specific regions. To determine these regions, let us consider the main case phase portrait in Figure 13c with a superposed on it "heat map" depicting $\gamma|\dot{s}|$ measured in the units of $\kappa$ in Figure 14.

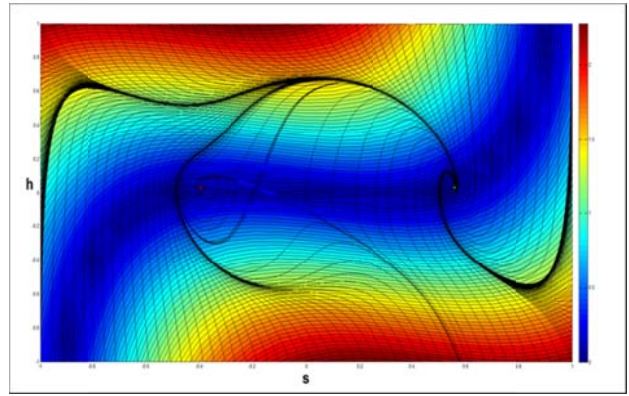

**Figure 14**: The main case phase portrait (Fig. 13c) with a superposed "heat map" showing the relative magnitudes of the feedback term $\gamma|\dot{s}|$ and the noise amplitude $\kappa$ ($\gamma|\dot{s}|$ measured in the units of $\kappa$). Deterministic effects are stronger in "hot" areas and weaker in "cold" areas.

It shows that the impact of noise is stronger near the turning points on trajectories (the dark blue strip in the figure) and weaker along the trajectories that correspond to large-scale, quasi-self-sustaining motion across the wells. Indeed, it can be expected that investor sentiment is more sensitive to news during the moments of indecisiveness in the market, when it is about to turn, and that positive news cannot easily reverse a market crash in full swing.

Let us now present the modeling results for system (13). We model $h(t), s(t)$ and $p(t)$ in the 1350 business-day period between 11/2004 and 04/2010 by substituting $\beta_1(\theta(t))$ with $\theta(t)$ empirically obtained for that period (Section 1.3.2) into equations (13).[31] This period has been

---

[31] We have shifted the empirically obtained $\beta_1(t)$ by -0.05 to increase the probability of market crash in simulations because, as has been previously discussed, a lower $\beta_1$ (higher $\theta$) reduces the barrier between the

chosen as it contains two characteristic regimes: the bull market (2005-2007) and the bear market (2008), which took place, respectively, in lower and higher temperature environments (Fig. 15).

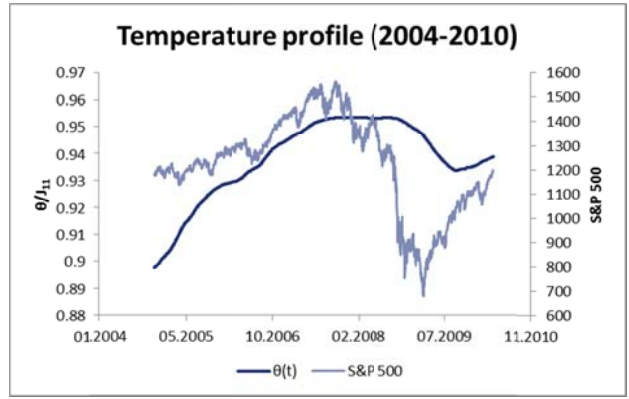

**Figure 15**: The empirically obtained temperature profile $\theta(t)$ to be applied for simulating stock market behavior.

Figure 16a depicts the average of 200 realizations of $s$, showing that sentiment on average declines from 2005 to 2008, reaches the minimum in 2008 and then rebounds between 2009 and 2010. This figure also plots a specific realization $s(t)$ that follows the above average pattern and so may be called a typical realization.

positive and negative halves of the potential well, making it easier for the sentiment to be moved across them. Further parameter values are reported in Figure 13c. Note that, as in the empirical study, $\beta_2$ is held constant for simplicity.

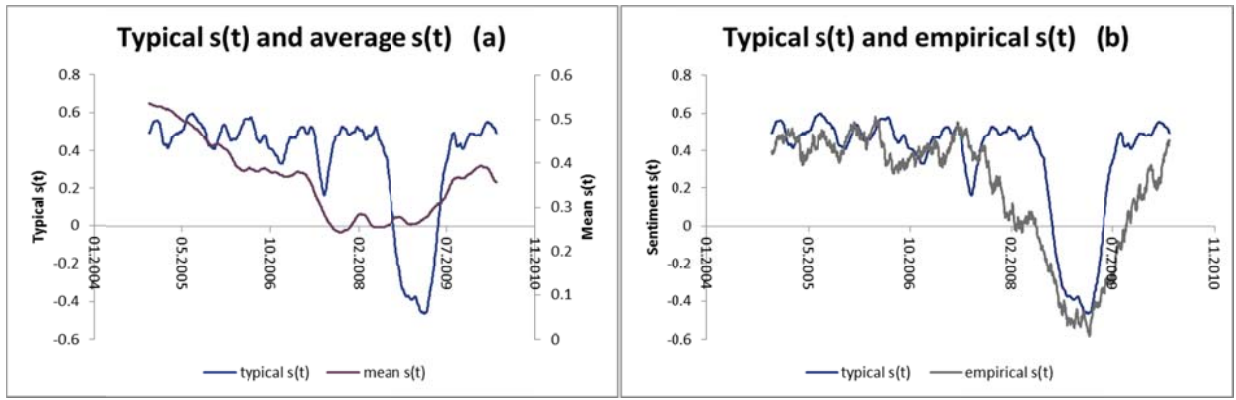

**Figure 16**: Simulated sentiment for the period 2004-2010. (a) A specific realization $s(t)$ vs. the average of 200 realizations plotted on a separate axis for improved visualization. (b) This same specific realization $s(t)$ and the empirically obtained $s(t)$ during that same period.

Let us now consider this specific realization in detail. During 2005-2007, $s$ moves around the equilibrium inside the positive well of the autonomous case (e.g. Fig. 13c,d). As temperature grows, the equilibrium shifts toward the origin and the amplitude of motion increases caused by the widening of the potential well. A further increase in temperature makes the well so wide as to allow $s$ to pass close to the trajectories that cross into the negative well, which increases the probability that the noise can force $s$ onto such trajectory. This event actually happens in 2008. Figure 16b compares the evolution of the simulated $s(t)$ with that of the empirically obtained $s(t)$ over the same period. It is evident that the simulated and empirical behaviors are similar.

Based on the typical realization of $s(t)$ in Figure 16, we construct the model price $p(t)$ from equation (13a), in which the values of the coefficients $a_1$ and $a_2$ have been taken from the empirical study (Table I) and $s_* = 0.35$ (the average empirical $s_*$ on this interval is 0.22), and compare $p(t)$ with the S&P 500 Index for that same period. Figure 17 shows the substantially similar behaviors of the simulated model price and the actual market price.

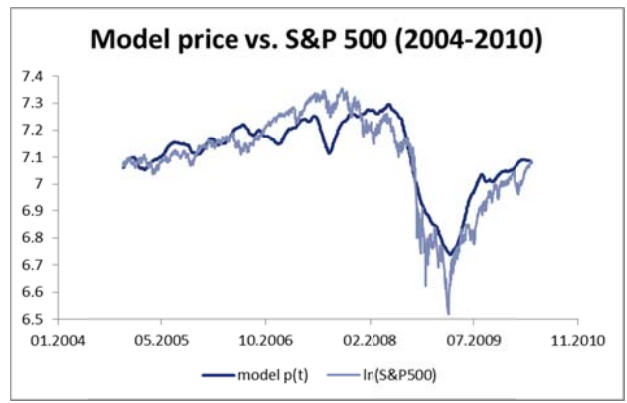

**Figure 17**: The evolution of the model price $p(t)$ obtained from the realization $s(t)$ in Fig. 16 and the evolution of the S&P 500 log price (2004-2010).

The evolution of sentiment and, therefore, of price is significantly modulated by the pattern of temperature variation. As the temperature profile is the only time-dependent factor that we have used to connect the model to observations, it is reasonable to conclude that temperature plays an important role in the formation of stock market regimes and, in particular, in the development of market rallies and crashes.

Another important aspect that warrants a separate discussion is the distribution of returns simulated by model (13). It would be interesting to learn whether the main characteristics of actual return distributions, such as positive kurtosis, heavy tails and nonzero skewness, can be explained by the model. As follows from equations (13), the term $\gamma \dot{s}$ in equation (13c) can cause strong changes in $s$, especially along the trajectories that correspond to quasi-self-sustaining motion across the wells, thus contributing to the events in the distribution tails. Also, the general asymmetry of the phase portrait (Fig. 13c,d) and the presence of the term $a_2(s - s_*)$ in equation (13a) can lead to a skewed distribution of returns.

Figure 18 shows the distributions of monthly returns generated by the theoretical model, the empirical model (based on the measured $H(t)$) and the S&P 500 Index. Both the theoretical and the empirical model return distributions resemble the distribution of actual returns, including the heavy tails. Note, however, that the tails of the theoretical and empirical distributions are truncated at relatively large deviations from the mean. This is most probably a consequence of the simplifying assumptions (including the approximation of the all-to-all interaction pattern) utilized in the models and, in the case of the empirical model, a result of measurement errors. It would therefore be incorrect to calculate kurtosis, skewness and other statistics of the model return distributions, as they evidently do not comprise the complete range of returns but only a portion of the range due to the cutoffs.[32] However, in the relevant range of returns, all three distributions exhibit similar shapes, as can be seen from the graphs which demonstrate visually similar "peakedness", skewness as well as the existence of (truncated) fat tails.

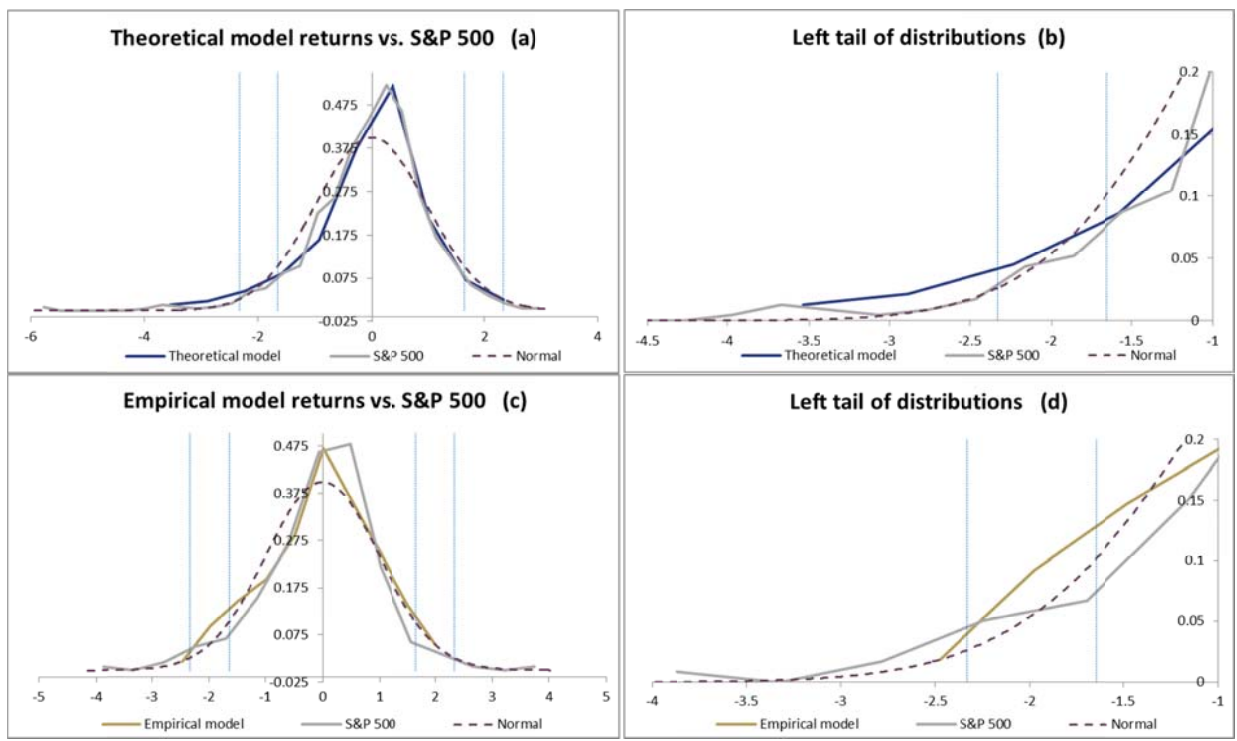

---

[32] Notwithstanding the truncated tails, the theoretical model distribution has (excess) kurtosis of 1.27 and skewness of -0.71; the corresponding values for the S&P 500 Index are 3.73 and -0.96, respectively.

Page 56 of 97

**Figure 18**: Distributions of 21-day log returns. (a,b) Theoretical model (60 years, i.e. 720 data points) and the S&P 500 Index (1950-2012). (c,d) Empirical model (1996-2012) and the S&P 500 Index (1996-2012). To make comparison easier, the distributions are normalized to yield zero mean and unit variance. The corresponding normal distributions are plotted as well. The horizontal axis, denominated in the units of standard deviation, indicates the normal distribution 1st and 5th percentiles.

The results presented in this section enable us to conclude that, first, there exist substantial similarities between the behavior of investor sentiment produced by the theoretical model and the empirical model; second, there exist substantial similarities in the price behavior generated by the theoretical model and that observed in the stock market; and, third, the theoretical model replicates characteristic features of the stock market, such as non-normally distributed returns.

## 2.2.3. Discussion

Results produced by model (13) are in agreement with the empirical study. In particular, the model exhibits two distinct behaviors represented by different characteristic frequencies at which investor sentiment evolves: mid-term oscillations in the positive well consistent with mid-term market trends and large-scale, quasi-self-sustaining motion consistent with market rallies and crashes.

Exogenous news flow is a random force that thrusts sentiment inside the potential well. It can occasionally force sentiment across the regions of different behavior, resulting in a switch of market regimes. We have seen that the influence of news is not uniform. There is a wide range of regimes between two extremes: the first where the news (noise) dominate leading to a nearly random price walk and the second where the news (noise) have low impact resulting in an almost deterministic price development. This implies that, according to the model, the market's evolution path can take it through a variety of regimes, some of which may have elements of deterministic dynamics.

We have demonstrated that a stochastically forced, deterministic nonlinear model described by equations (13) is capable of generating complex behaviors representative of actual market regimes. For example, we have been able to approximately reproduce the behavior of the S&P 500 Index from 2004 to 2010 by substituting the empirical temperature profile for that same period into the model (which implies that temperature plays a significant role in the development of market regimes). The model has been shown to replicate essential features of the actual distributions of stock market returns, such as nonzero skewness, positive kurtosis and fat (albeit truncated) tails. In addition, the model has provided an explanation of market trends and crashes in terms of characteristic frequencies of sentiment development.

The appearance of characteristic frequencies in the model is intriguing and deserves conjecture on its causes and implications. Model (13) is a heavily approximated version of the more general Ising model defined by equation (A1) in Appendix A, the dynamics of which are largely determined by the topology of interaction among the agents that constitute it. In particular, model (13) has been derived under the assumption that the pattern of interaction in equation (A1) is all to all. It is known that the all-to-all pattern prevents Ising spins (agents) from assembling into clusters, whereas interactions that are not as radically simplified can produce heterogeneous structures (e.g. Cont and Bouchaud (2000)) – which, in our case, represent the clusters of investors and analysts characterized by coaligned sentiments.[33]

---

[33] As mentioned earlier, the all-to-all topology is the leading-order approximation for general interaction topology in the Ising model and so is a convenient starting point for studying its key features. We note that prices generated by the empirical model (eq. 4 and 10), which also employs the all-to-all interaction pattern, can contain indirect evidence of cluster dynamics because the input variable $H(t)$ is based on empirical data.

The presence of clusters can lead to diverse dynamics of interactions on various timescales. This is because a cluster's size determines its reaction time to various disturbances, such as adjacent clusters, random influences or external forces (the news), such that the larger the size, the slower the reaction. As different reaction times of differently-sized clusters to incoming information can represent investment horizons or allocation processes of various investor groups, the general model (eq. A1) can simulate the dynamics of market participants ranging from private investors to pension plans.

Consequently, the general model (eq. A1) can exhibit multiple frequencies of interaction. We may expect that the actual stock market also contains different characteristic frequencies, each of which signifies a distinct behavior that together, through interactions, produce a constantly shifting pattern of regimes driving the development of market price. And indeed, the fact that a drastically simplified market model captured two characteristic frequencies that correspond to actual market regimes supports this expectation.[34]

We would like to conclude by returning to the discussion of the possible predictability of market returns that we have touched upon in the beginning of this section. Recall that the characteristic frequencies emerge in the model as an inertial effect caused by the feedback relation between direct information and price. The existence of inertia has another consequence: the model acquires a resistance to change in direction and, as a result, its behavior becomes predictable in situations where inertial effects outweigh noise. In other words, based on the theoretical results and the support of the empirical evidence, we can expect the stock market to include regimes with some elements of deterministic dynamics, which can in principle be predicted.

---

[34] The arguably quasiperiodic pattern of the autocorrelations of $H(t)$ (Fig. 11a) can be a further sign of the presence of characteristic frequencies in the market price series.

As an illustration, let us consider how the models developed here could be applied for making a market forecast. A possible procedure is as follows: measure daily $H(t)$, obtain $s(t)$ from equation (4), find the corresponding location $h(t), s(t)$ on the phase portrait (e.g. Fig. 13c) and forecast the following day's return expectation which, being a function of the current day's location on the phase portrait, may be nonzero. Of course, model (13) is likely too crude to produce a meaningful forecast. More sophisticated patterns of interaction may be needed to construct more realistic models. In the light of the above discussion, such randomly-driven deterministic models are likely to generate complex dynamics, resulting in the presence of random, deterministic chaotic and deterministic non-chaotic behaviors. However, unlike the case of the random walk, forecasting the behavior of complex deterministic systems can be possible by blending models with observations, as is done, for example, in meteorology and oceanography and can probably be done as well in the case of the financial markets.[35]

## 3. Conclusion

This paper introduced a framework for understanding stock market behavior, upon which the model of stock market dynamics was developed and studied. We have demonstrated, empirically and/or theoretically, that according to this model:

1. The stock market dynamics can be explained in terms of the interaction among three variables: direct information (defined in Section 1.1), investor sentiment (defined in Section 1.2) and market price – as the theoretical and empirical results are in agreement and show a reasonably good fit with the observations.

---

[35] This is the direction of our future research. As a preliminary finding, we would like to report that simple prototypes of algorithmic trading strategies built upon the framework developed here have produced promising backtested results, supportive of the above-described approach to market forecasting.

2. The effectiveness of the influence of information on investors is determined by the degree of directness of its interpretation in relation to the expectations of future market performance. Direct information – information that explicitly mentions the direction of anticipated market movement – impacts investors most.

3. In addition to being forced by direct information, the evolution of investor sentiment is also significantly influenced, first, by the interaction among investors and, second, by the level of disorder in the market given by the vacillating balance between the herding and random behavior of investors. The influence due to the second factor is determined by the economic analog of temperature (introduced in Section 1.2.1), the profile of which can be deduced from the empirical model.

4. Market price develops differently on different timescales: on shorter timescales (e.g. days and weeks), price changes proportionally to the change in sentiment, while on longer timescales (e.g. months and years), price changes proportionally to the deviation of sentiment from a reference level that is generally nonzero. As a result, price evolution is naturally decomposed into a long-term, large-scale variation and short-term, small-to-mid-scale fluctuations.

5. In the market, herding behavior prevailed (to a small degree) over random behavior during the studied period (1996-2012). However, the level at which these behaviors were balanced varied gradually over time such that, generally, herding increased during bull markets and decreased during bear markets. The level of balance affects market volatility and the probability of market crashes. There is a connection between changes in this balance and observed economic fluctuations (the business cycle).

6. Long-term sentiment trends show a substantial temporal lead over long-term market price trends. This result may be of practical importance for the development of trend-following strategies, as a change in sentiment trend could be a precursor to a change in price trend.

7. Information related to price changes plays an important role in market dynamics by inducing a feedback loop: information -> sentiment -> price -> information. This leads to a nonlinear dynamic of interaction, which explains the existence of certain market behaviors, such as trends, rallies and crashes. It is responsible for the familiar non-normal shape of the stock market return distribution.

8. Exogenous news act as a random force that displaces investor sentiment from the equilibrium and occasionally causes market dynamics to switch from one regime to another. Two equilibrium states are possible. In the first regime, sentiment and price remain constant. In the second regime, sentiment oscillates between extreme negative and positive values, driving alternating bear and bull markets.

9. The coupling of price and information creates feedback, whereby information causes price changes and price changes generate information (paragraph 7). Additionally, the resulting dynamic further complicates the picture, leading to a delayed reaction between these variables, such that, for example, today's price change may be influenced by information related to past price changes, along with other news, over previous days, weeks and months. This implies that the notion of cause and effect does not simplistically apply to market dynamics, as cause and effect become, in a sense, intertwined.

10. The nonlinear dynamic underlying sentiment evolution contains both deterministic and random components. The deterministic component in some market regimes is stronger, while other regimes may be dominated by the random component. Developing an understanding of which market regime takes place at which time can be possible by combining models and observations, which could potentially facilitate market return forecasts.

## Acknowledgements

We are grateful to Dow Jones & Company for providing access to Factiva.com news archive. We would also like to thank John Orthwein for editing this paper and contributing ideas on its readability.

## Appendix A: Equation for sentiment dynamics

We base the derivation of a dynamic equation for sentiment evolution in the model studied in Section 2 on a physical analogy in which two sets of interacting Ising spins $s_i = \pm 1, i = 1 \ldots N_s$ and $h_j = \pm 1, j = 1 \ldots N_h$ are acted upon by external magnetic fields $b_s(t)$ and $b_h(t)$, respectively. Additionally, this analysis yields, as a particular case, a dynamic equation for a single set of Ising spins $s_i = \pm 1, i = 1 \ldots N_s$ in the presence of an external magnetic field $H(t)$, which is a physical analogy of the model studied in Section 1. This particular case is similar to the statistical mechanics problem treated by Glauber (1963), Suzuki and Kubo (1968), and Ovchinnikov and Onishchuk (1988).

### A1. Properties of the two-component Ising system

The Hamiltonian[36] of the Ising system has the form:

$$\mathcal{H} = -\frac{1}{2}\sum_{i \neq k}^{N_s} J_{s_{ik}} s_i s_k - \frac{1}{2}\sum_{i \neq k}^{N_h} J_{h_{ik}} h_i h_k - \sum_{i,j}^{N_s N_h} J_{sh_{ij}} s_i h_j - \mu_s \sum_i^{N_s} b_{s_i}(t) s_i - \mu_h \sum_i^{N_h} b_{h_i}(t) h_i,$$

where $J_{s_{ik}}, J_{h_{ik}}$ and $J_{sh_{ij}}$ are coefficients that determine the strength of interaction among spins; $\mu_s$ and $\mu_h$ are magnetic moments per spin; and $b_{s_i}(t)$ and $b_{h_i}(t)$ are external magnetic fields.

If we assume that

$J_{s_{ik}} = J_s = J_s(N_s),$

$J_{h_{ik}} = J_h = J_h(N_h),$

$J_{sh_{ij}} = J_{sh} = J_{sh}(N_s, N_h),$

---

[36] In this appendix we revert to the notation and terminology commonly used in physics.

$b_{s_i}(t) = b_s(t),$

$b_{h_i}(t) = b_h(t),$

then the Hamiltonian becomes

$$\mathcal{H}(S, H, t) = -\left[\frac{J_s}{2}(S^2 - N_s) + \frac{J_h}{2}(H^2 - N_h) + J_{sh}SH + \mu_s b_s(t)S + \mu_h b_h(t)H\right], \quad (A1)$$

where

$S = \sum_{i=1}^{N_s} s_i, \quad H = \sum_{j=1}^{N_h} h_j.$

In what follows we assume that in the thermodynamic limit ($N_s \to \infty, N_h \to \infty$) the values of $J_s N_s$, $J_h N_h$, $J_{sh} N_s$ and $J_{sh} N_h$ are finite and denote them as

$$\lim_{N_s \to \infty}(J_s N_s) = J_{11}, \quad \lim_{N_h \to \infty}(J_h N_h) = J_{22}, \quad \lim_{\substack{N_s \to \infty \\ N_h \to \infty}}(J_{sh} N_h) = J_{12}, \quad \lim_{\substack{N_s \to \infty \\ N_h \to \infty}}(J_{sh} N_s) = J_{21}. \quad (A2)$$

In situations where the external fields are stationary, $b_s(t) = b_s$ and $b_h(t) = b_h$, the Ising system can be in the state of thermodynamic equilibrium. In this case, the probability of finding it in the state with the values of total spins equal to $S$ and $H$, respectively, is given by

$$P_0(S, H) = \frac{g(S)g(H)e^{-\frac{E(S,H)}{\theta}}}{Z}, \quad (A3)$$

where $E(S, H) = \mathcal{H}(S, H)$ is the energy of the system and $\theta$ is temperature measured in the units of energy. The functions $g(S)$ and $g(H)$, given by

$$g(S) = \frac{N_s!}{\left(\frac{N_s + S}{2}\right)!\left(\frac{N_s - S}{2}\right)!} \quad \text{and} \quad g(H) = \frac{N_h!}{\left(\frac{N_h + H}{2}\right)!\left(\frac{N_h - H}{2}\right)!},$$

are the degrees of degeneracy of the states for which total spins are equal to $S$ and $H$, respectively. $Z$ is a normalization factor determined from the condition

$$\sum_{S=-N_s}^{N_s} \sum_{H=-N_h}^{N_h} P_0(S,H) = 1,$$

which leads to

$$Z(N_s, N_h, t) = \sum_{S=-N_s}^{N_s} \sum_{H=-N_h}^{N_h} g(S)g(H) e^{-\frac{E(S,H)}{\theta}},$$

called the partition function.

In the limit of large $N_s$ and $N_h$ ($N_s \gg 1, N_h \gg 1$), it is convenient to change from discrete variables $S$ ($-N_s, -N_s + 2, \ldots N_s - 2, N_s$) and $H$ ($-N_h, -N_h + 2, \ldots N_h - 2, N_h$) to quasi-continuous variables

$$s = \frac{S}{N_s}, \quad (-1 \leq s \leq 1),$$

$$h = \frac{H}{N_h}, \quad (-1 \leq h \leq 1).$$

We can rewrite the above equations in new variables $s$ and $h$ to obtain

$$E(s,h) = -\frac{N_s}{2}\left(J_{11}s^2 + J_{12}sh + 2\mu_s b_s s - \frac{J_{11}}{N_s}\right) - \frac{N_h}{2}\left(J_{22}h^2 + J_{21}sh + 2\mu_h b_h h - \frac{J_{22}}{N_h}\right),$$

$$\mathfrak{S}(s,h) = \ln g(S)g(H)$$

$$= \frac{N_s}{2}\left(s \ln\left(\frac{1-s}{1+s}\right) - \ln\frac{(1-s^2)}{4} + \frac{1}{N_s}\ln\left(\frac{2}{\pi N_s(1-s^2)}\right)\right)$$

$$+ \frac{N_h}{2}\left(h \ln\left(\frac{1-h}{1+h}\right) - \ln\frac{(1-h^2)}{4} + \frac{1}{N_h}\ln\left(\frac{2}{\pi N_h(1-h^2)}\right)\right).$$

Then the equilibrium distribution function $P_0(s,h)$ takes the form:

$$P_0(s,h) = \frac{\exp\left\{\mathfrak{S}(s,h) - \frac{E(s,h)}{\theta}\right\}}{Z} = \frac{\exp\left\{-\frac{F(s,h)}{\theta}\right\}}{Z},$$

where

$$F(s,h) = E(s,h) - \theta\mathfrak{S}(s,h)$$

$$= -\frac{N_s}{2}\left(J_{11}s^2 + J_{12}sh + 2\mu_s b_s s + \theta\left(s\ln\left(\frac{1-s}{1+s}\right) - \ln\frac{(1-s^2)}{4}\right) + O\left(\frac{1}{N_s}\right)\right)$$

$$- \frac{N_h}{2}\left(J_{22}h^2 + J_{21}sh + 2\mu_h b_h h + \theta\left(h\ln\left(\frac{1-h}{1+h}\right) - \ln\frac{(1-h^2)}{4}\right) + O\left(\frac{1}{N_h}\right)\right).$$

Let us find the values of $s$ and $h$ for which the distribution function $P_0(s,h)$ has a maximum (minimum). Neglecting terms $O\left(\frac{1}{N_s}\right)$ and $O\left(\frac{1}{N_h}\right)$ in $F(s,h)$, we find from the extremum condition $\frac{\partial F(s,h)}{\partial s} = 0$ and $\frac{\partial F(s,h)}{\partial h} = 0$ that

$$\begin{cases} 2J_{11}s + \left(J_{12} + \frac{N_h}{N_s}J_{21}\right)h + 2\mu_s b_s = \theta\ln\left(\frac{1+s}{1-s}\right), \\ 2J_{22}h + \left(J_{21} + \frac{N_s}{N_h}J_{12}\right)s + 2\mu_h b_h = \theta\ln\left(\frac{1+h}{1-h}\right). \end{cases}$$

Accounting for $\frac{J_{21}}{J_{12}} = \frac{N_s}{N_h}$, we can rewrite this system of equations as

$$\begin{cases} J_{11}s + J_{12}h + \mu_s b_s = \frac{\theta}{2}\ln\left(\frac{1+s}{1-s}\right), \\ J_{22}h + J_{21}s + \mu_h b_h = \frac{\theta}{2}\ln\left(\frac{1+h}{1-h}\right), \end{cases}$$

or equivalently:

$$\begin{cases} s = \tanh(\beta(J_{11}s + J_{12}h + \mu_s b_s)), \\ h = \tanh(\beta(J_{22}h + J_{21}s + \mu_h b_h)), \end{cases} \tag{A4}$$

where $\beta = 1/\theta$.

As a simple example, let us consider a case in which $b_s = b_h = 0, J_{11} = J_{22} = J_{12} = J_{21} = J_0$. Then $(A4)$ is reduced to

$$\begin{cases} s = \tanh(j_0(s+h)), \\ h = \tanh(j_0(s+h)), \end{cases}$$

where $j_0 = \beta J_0$. It follows that $s = h$ and $s = \tanh(2j_0 s)$. If $2j_0 \leq 1$, there is only one solution: $s = h = 0$ (zero magnetization). If $2j_0 > 1$, there exist three solutions with $s = 0, \pm s_0$, two of which, namely $s = \pm s_0$ (nonzero magnetization), correspond to the maxima of the distribution function $P_0(s, h)$. Thus, the critical temperature of the phase transition is in this case $\theta_c = 2J_0$. We note that a more general case is treated in Appendix C.

It is worthwhile noting that phase transition is possible even in the case of $J_{11} = J_{22} = 0$. Then:

$$\begin{cases} s = \tanh(j_{12} h), \\ h = \tanh(j_{21} s), \end{cases}$$

where $j_{12} = \beta J_{12}, j_{21} = \beta J_{21}$, which yields nonzero solutions for $j_{12} j_{21} > 1$.

## A2. Master (kinetic) equation

To describe non-equilibrium dynamics, we introduce distribution function $P(S, H, t)$ that defines the probability that the values of the total spins of configurations $\{s_i\}$ and $\{h_i\}$ at time $t$ are equal to $S$ and $H$, respectively. If $P(S, H, t)$ is known, we can calculate any statistical quantity at any moment $t$. For instance, the expectation values of total spins $S$ and $H$ are given by

$$S(t) = \langle S \rangle_t = \sum_{H=-N_h}^{N_h} \sum_{S=-N_s}^{N_s} S\, P(S, H, t), \tag{A5a}$$

$$H(t) = \langle H \rangle_t = \sum_{S=-N_s}^{N_s} \sum_{H=-N_h}^{N_h} H\, P(S,H,t). \tag{A5b}$$

Note that summation in (A5) is carried out across the integer values: $S = -N_s, -N_s + 2, \ldots N_s - 2, N_s$ and $H = -N_h, -N_h + 2, \ldots N_h - 2, N_h$. We can obtain $P(S,H,t)$ by deriving and solving the corresponding master equation for the evolution of $P(S,H,t)$. To formulate the master equation, we define the probability of the system transiting from one configuration of spins into another per unit time as $W(S,H;\,S',\,H')$, where it is assumed that transitions occur when spins spontaneously flip as a result of uncontrolled energy exchange with a heat bath. Considering that $P(S,H,t)$ evolves only due to these transitions, we conclude that it must satisfy the following master equation:

$$\frac{d\,P(S,H,t)}{dt} = -\sum_{S'=-N_s}^{N_s} \sum_{H'=-N_h}^{N_h} W(S,H;\,S',H')\,P(S,H,t)$$

$$+ \sum_{S'=-N_s}^{N_s} \sum_{H'=-N_h}^{N_h} P(S',H',t)W(S',H';S,H). \tag{A6}$$

Note that the first sum in ($A6$) represents the total probability per unit time that the system 'flips out' of the state $\{S,H\}$, whereas the second sum corresponds to the total probability per unit time that the system 'flips in' to the state $\{S,H\}$.

Let us assume that transitions occur only due to single spin flips, i.e. $S$ and $H$ change at each transition by $\pm 2$. Then equation ($A6$) becomes

$$\frac{d\,P(S,H,t)}{dt} = -\big(W(S,H;\,S+2,H) + W(S,H;\,S-2,H) + W(S,H;\,S,H+2)$$

$$+ W(S,H;\,S,H-2)\big)P(S,H,t) + W(S+2,H;\,S,H)P(S+2,H,t)$$

$$+ W(S-2,H;\,S,H)P(S-2,H,t) + W(S,H+2;\,S,H)P(S,H+2,t)$$

$$+ W(S,H-2;\,S,H)P(S,H-2,t). \tag{A7}$$

Therefore, the problem is now reduced to, first, deducing the form of $W$ and then solving $(A7)$. To determine $W$ we can use the detailed balance condition:

$$W(S,H; S',H')P_0(S,H) = W(S',H'; S,H)P_0(S',H').$$

Here $P_0(S,H)$ is the equilibrium distribution function given by $(A3)$. When the detailed balance condition is satisfied, the number of transitions per unit time in the thermodynamic equilibrium state from any configuration $\{S,H\}$ into any other configuration $\{S',H'\}$ exactly equals the number of transitions in the opposite direction. This condition imposes certain constraints on the probability $W$, as it must satisfy

$$\frac{W(S,H; S',H')}{W(S',H'; S,H)} = \frac{P_0(S',H')}{P_0(S,H)} = \frac{g(S')g(H')}{g(S)g(H)} \exp\left\{-\frac{E(S',H') - E(S,H)}{\theta}\right\}. \tag{A8}$$

According to $(A8)$, the probabilities $W$ in $(A7)$ satisfy the following equations:

$$\begin{cases} \dfrac{W(S,H; S\pm 2, H)}{W(S\pm 2, H; S,H)} = \dfrac{P_0(S\pm 2,H)}{P_0(S,H)} = \dfrac{g(S\pm 2)}{g(S)} \exp\left\{-\dfrac{E(S\pm 2,H) - E(S,H)}{\theta}\right\}, \\ \dfrac{W(S,H; S, H\pm 2)}{W(S, H\pm 2; S,H)} = \dfrac{P_0(S,H\pm 2)}{P_0(S,H)} = \dfrac{g(H\pm 2)}{g(H)} \exp\left\{-\dfrac{E(S,H\pm 2) - E(S,H)}{\theta}\right\}. \end{cases}$$

Using expressions for $g(S)$, $g(H)$ and $E(S,H)$, we obtain

$$\begin{cases} \dfrac{W(S,H; S\pm 2, H)}{W(S\pm 2, H; S,H)} = \dfrac{\dfrac{(N_s \mp S)}{2}}{\left(\dfrac{(N_s \pm S)}{2} + 1\right)} \exp\{\pm 2\beta(J_s(S\pm 1) + J_{sh}H + \mu_s b_s)\}, \\ \dfrac{W(S,H; S, H\pm 2)}{W(S, H\pm 2; S,H)} = \dfrac{\dfrac{(N_h \mp H)}{2}}{\left(\dfrac{(N_h \pm H)}{2} + 1\right)} \exp\{\pm 2\beta(J_h(H\pm 1) + J_{sh}S + \mu_h b_h)\}. \end{cases} \tag{A9}$$

We seek $W(S,H; S',H')$ that satisfy $(A9)$ in the form:

$$W(S,H; S\pm 2, H) = \frac{(N_s \mp S)}{2} \frac{w_s}{(1 + \exp\{\mp 2\beta(J_s(S\pm 1) + J_{sh}H + \mu_s b_s)\})},$$

$$W(S \pm 2, H; S, H) = \left(\frac{(N_s \pm S)}{2} + 1\right) \frac{w_s}{(1 + \exp\{\pm 2\beta(J_s(S \pm 1) + J_{sh}H + \mu_s b_s)\})},$$

$$W(S, H; S, H \pm 2) = \frac{(N_h \mp H)}{2} \frac{w_h}{(1 + \exp\{\mp 2\beta(J_h(H \pm 1) + J_{sh}S + \mu_h b_h)\})},$$

$$W(S, H \pm 2; S, H) = \left(\frac{(N_h \pm H)}{2} + 1\right) \frac{w_h}{(1 + \exp\{\pm 2\beta(J_h(H \pm 1) + J_{sh}S + \mu_h b_h)\})},$$

where $w_s$ and $w_h$ are the corresponding relaxation rates and $\beta = 1/\theta$, as defined before.

Let us define for convenience

$$W(S, H; S \pm 2, H) = \frac{(N_s \mp S)}{2} w(S, H; S \pm 2, H),$$

$$W(S \pm 2, H; S, H) = \left(\frac{(N_s \pm S)}{2} + 1\right) w(S \pm 2, H; S, H),$$

$$W(S, H; S, H \pm 2) = \frac{(N_h \mp H)}{2} w(S, H; S, H \pm 2),$$

$$W(S, H \pm 2; S, H) = \left(\frac{(N_h \pm H)}{2} + 1\right) w(S, H \pm 2; S, H),$$

where

$$w(S, H; S \pm 2, H) = \frac{w_s}{(1 + \exp\{\mp 2\beta(J_s(S \pm 1) + J_{sh}H + \mu_s b_s)\})},$$

$$w(S \pm 2, H; S, H) = \frac{w_s}{(1 + \exp\{\pm 2\beta(J_s(S \pm 1) + J_{sh}H + \mu_s b_s)\})},$$

$$w(S, H; S, H \pm 2) = \frac{w_h}{(1 + \exp\{\mp 2\beta(J_h(H \pm 1) + J_{sh}S + \mu_h b_h)\})},$$

$$w(S, H \pm 2; S, H) = \frac{w_h}{(1 + \exp\{\pm 2\beta(J_h(H \pm 1) + J_{sh}S + \mu_h b_h)\})}.$$

Then system (A8) takes the final form:

$$\frac{d\,P(S,H,t)}{dt} = -\left(\frac{(N_s+S)}{2}w(S,H;\,S-2,H) + \frac{(N_s-S)}{2}w(S,H;\,S+2,H)\right.$$

$$\left. + \frac{(N_h+H)}{2}w(S,H;\,S,H-2) + \frac{(N_h-H)}{2}w(S,H;\,S,H+2)\right)P(S,H,t)$$

$$+ \left(\frac{(N_s+S)}{2}+1\right)w(S+2,H;S,H)P(S+2,H,t)$$

$$+ \left(\frac{(N_s-S)}{2}+1\right)w(S-2,H;S,H)P(S-2,H,t)$$

$$+ \left(\frac{(N_h+H)}{2}+1\right)w(S,H+2;S,H)P(S,H+2,t)$$

$$+ \left(\frac{(N_h-H)}{2}+1\right)w(S,H-2;S,H)P(S,H-2,t). \qquad (A10)$$

For solving $(A10)$ it is necessary to define the initial distribution $P(S,H,0) = f_0(S,H)$, as well as the boundary conditions, which in this case are $P(S,H,t) \equiv 0$ for $|S| > N_s$ and $|H| > N_h$. We also note that equations $(A10)$ are applicable in the case of nonstationary external fields $b_s(t)$ and $b_h(t)$.

## A3. Dynamic equations for average spins

We can derive the equations of motion for average spins $S(t) = \langle S \rangle_t$ and $H(t) = \langle H \rangle_t$ by differentiating equation $(A5)$ with respect to time and using equations $(A10)$:

$$\frac{d\,S(t)}{dt} = \sum_{S=-N_s}^{N_s}\sum_{H=-N_h}^{N_h} S\,\frac{d\,P(S,H,t)}{dt} = \sum_{n=1}^{8} S_n(t),$$

$$\frac{d\,H(t)}{dt} = \sum_{S=-N_s}^{N_s}\sum_{H=-N_h}^{N_h} H\,\frac{d\,P(S,H,t)}{dt} = \sum_{n=1}^{8} H_n(t),$$

where functions $S_n(t)$ and $H_n(t)$ represent the results of summation of the $n$-th term in the r.h.s. of ($A10$) multiplied by $S$ and $H$, respectively.

We obtain the following expressions for $S_n(t)$ and $H_n(t)$:

$$\begin{pmatrix} S_1(t) \\ H_1(t) \end{pmatrix} = -\sum_{S=-N_s}^{N_s} \sum_{H=-N_h}^{N_h} \begin{pmatrix} S \\ H \end{pmatrix} \frac{(N_s + S)}{2} w(S,H; S-2, H) P(S,H,t)$$

$$= -\frac{1}{2} \langle \begin{pmatrix} S \\ H \end{pmatrix} (N_s + S) w(S,H; S-2, H) \rangle_t,$$

$$\begin{pmatrix} S_2(t) \\ H_2(t) \end{pmatrix} = -\sum_{S=-N_s}^{N_s} \sum_{H=-N_h}^{N_h} \begin{pmatrix} S \\ H \end{pmatrix} \frac{(N_s - S)}{2} w(S,H; S+2, H) P(S,H,t)$$

$$= -\frac{1}{2} \langle \begin{pmatrix} S \\ H \end{pmatrix} (N_s - S) w(S,H; S+2, H) \rangle_t,$$

$$\begin{pmatrix} S_3(t) \\ H_3(t) \end{pmatrix} = -\sum_{S=-N_s}^{N_s} \sum_{H=-N_h}^{N_h} \begin{pmatrix} S \\ H \end{pmatrix} \frac{(N_h + H)}{2} w(S,H; S, H-2) P(S,H,t)$$

$$= -\frac{1}{2} \langle \begin{pmatrix} S \\ H \end{pmatrix} (N_h + H) w(S,H; S, H-2) \rangle_t,$$

$$\begin{pmatrix} S_4(t) \\ H_4(t) \end{pmatrix} = -\sum_{S=-N_s}^{N_s} \sum_{H=-N_h}^{N_h} \begin{pmatrix} S \\ H \end{pmatrix} \frac{(N_h - H)}{2} w(S,H; S, H+2) P(S,H,t)$$

$$= -\frac{1}{2} \langle \begin{pmatrix} S \\ H \end{pmatrix} (N_h - H) w(S,H; S, H+2) \rangle_t,$$

$$\begin{pmatrix} S_5(t) \\ H_5(t) \end{pmatrix} = \sum_{S=-N_s}^{N_s} \sum_{H=-N_h}^{N_h} \binom{S}{H} \frac{(N_s + S + 2)}{2} w(S+2, H; S, H) P(S+2, H, t)$$

$$= \frac{1}{2} \sum_{S'=-N_s+2}^{N_s+2} \sum_{H=-N_h}^{N_h} \binom{S'-2}{H} (N_s + S') w(S', H; S'-2, H) P(S', H, t)$$

$$= \frac{1}{2} \sum_{S'=-N_s}^{N_s} \sum_{H=-N_h}^{N_h} \binom{S'-2}{H} (N_s + S') w(S', H; S'-2, H) P(S', H, t)$$

$$= \frac{1}{2} \left\langle \binom{S-2}{H} (N_s + S) w(S, H; S-2, H) \right\rangle_t.$$

To derive this last expression, we, having changed variables $S' = S + 2$, omitted the term with $S' = N_s + 2$ as $P(N_s + 2, H, t) \equiv 0$ and added the term with $S' = -N_s$ as the summated function contains the multiplier $(N_s + S')$ that vanishes at such value of $S'$, which has enabled us to represent this expression in canonical form above. Similarly,

$$\begin{pmatrix} S_6(t) \\ H_6(t) \end{pmatrix} = \sum_{S=-N_s}^{N_s} \sum_{H=-N_h}^{N_h} \binom{S}{H} \frac{(N_s - S + 2)}{2} w(S-2, H; S, H) P(S-2, H, t)$$

$$= \frac{1}{2} \sum_{S'=-N_s-2}^{N_s-2} \sum_{H=-N_h}^{N_h} \binom{S'+2}{H} (N_s - S') w(S', H; S'+2, H) P(S', H, t)$$

$$= \frac{1}{2} \sum_{S'=-N_s}^{N_s} \sum_{H=-N_h}^{N_h} \binom{S'+2}{H} (N_s - S') w(S', H; S'+2, H) P(S', H, t)$$

$$= \frac{1}{2} \left\langle \binom{S+2}{H} (N_s - S) w(S, H; S+2, H) \right\rangle_t,$$

$$\binom{S_7(t)}{H_7(t)} = \sum_{S=-N_s}^{N_s} \sum_{H=-N_h}^{N_h} \binom{S}{H} \frac{(N_h + H + 2)}{2} w(S, H+2;\, S, H) P(S, H+2, t)$$

$$= \frac{1}{2} \sum_{S=-N_s}^{N_s} \sum_{H'=-N_h+2}^{N_h+2} \binom{S}{H'-2} (N_h + H') w(S, H';\, S, H'-2) P(S, H', t)$$

$$= \frac{1}{2} \sum_{S=-N_s}^{N_s} \sum_{H'=-N_h}^{N_h} \binom{S}{H'-2} (N_h + H') w(S, H';\, S, H'-2) P(S, H', t)$$

$$= \frac{1}{2} \left\langle \binom{S}{H-2} (N_h + H) w(S, H;\, S, H-2) \right\rangle_t,$$

$$\binom{S_8(t)}{H_8(t)} = \sum_{S=-N_s}^{N_s} \sum_{H=-N_h}^{N_h} \binom{S}{H} \frac{(N_h - H + 2)}{2} w(S, H-2;\, S, H) P(S, H-2, t)$$

$$= \frac{1}{2} \sum_{S=-N_s}^{N_s} \sum_{H'=-N_h-2}^{N_h-2} \binom{S}{H'+2} (N_h - H') w(S, H';\, S, H'+2) P(S, H', t)$$

$$= \frac{1}{2} \sum_{S=-N_s}^{N_s} \sum_{H'=-N_h}^{N_h} \binom{S}{H'+2} (N_h - H') w(S, H';\, S, H'+2) P(S, H', t)$$

$$= \frac{1}{2} \left\langle \binom{S}{H+2} (N_h - H) w(S, H;\, S, H+2) \right\rangle_t.$$

We can now sum the above expressions to obtain the following equations for $S(t)$ and $H(t)$:

$$\begin{cases} \dfrac{dS(t)}{dt} = \langle F_s(S, H) \rangle_t, \\ \dfrac{dH(t)}{dt} = \langle F_h(S, H) \rangle_t, \end{cases} \tag{A11}$$

where

$$F_s(S, H) = -(N_s + S) w(S, H;\, S-2, H) + (N_s - S) w(S, H;\, S+2, H),$$

$$F_h(S, H) = -(N_h + H) w(S, H;\, S, H-2) + (N_h - H) w(S, H;\, S, H+2).$$

We note that system $(A11)$ is not closed because $\langle F_s(S, H)\rangle_t$ and $\langle F_h(S, H)\rangle_t$ cannot in general be reduced to be functions of $S(t)$ and $H(t)$. However, it would be reasonable to assume that in the thermodynamic limit ($N_s \gg 1, N_h \gg 1$) fluctuations of total spins $S$ and $H$ around their instantaneous average values $S(t)$ and $H(t)$ are small in comparison with these same average values. This situation is analogous to that where a system is in thermodynamic equilibrium. We can then write

$$\langle F_s(S,H)\rangle_t = F_s(\langle S\rangle_t, \langle H\rangle_t) = F_s(S(t), H(t))$$

and

$$\langle F_h(S,H)\rangle_t = F_h(\langle S\rangle_t, \langle H\rangle_t) = F_h(S(t), H(t)).$$

We introduce new variables,

$$s(t) = \frac{\langle S\rangle_t}{N_s} = \frac{S(t)}{N_s},$$

$$h(t) = \frac{\langle H\rangle_t}{N_h} = \frac{H(t)}{N_h},$$

and express $F_s(S(t), H(t))$ and $F_h(S(t), H(t))$ via these variables to obtain in the limit $N_s \to \infty, N_h \to \infty$ the following equations:

$$F_s(S(t), H(t)) = N_s \left[ -(1+s(t))\frac{w_s}{1+\exp\{2\beta(J_{11}s(t)+J_{12}h(t)+\mu_s b_s(t))\}} \right.$$
$$\left. + (1-s(t))\frac{w_s}{1+\exp\{-2\beta(J_{11}s(t)+J_{12}h(t)+\mu_s b_s(t))\}} \right]$$
$$= N_s w_s \left[ \tanh\{\beta(J_{11}s(t)+J_{12}h(t)+\mu_s b_s(t))\} - s(t) \right]$$

and

$$F_h\big(S(t), H(t)\big) = N_h\left[-(1+h(t))\frac{w_h}{1+\exp\{2\beta(J_{22}h(t)+J_{21}s(t)+\mu_h b_h(t))\}}\right.$$

$$\left.+(1-h(t))\frac{w_h}{1+\exp\{-2\beta(J_{22}h(t)+J_{21}s(t)+\mu_h b_h(t))\}}\right]$$

$$= N_h w_h\big[\tanh\{\beta(J_{22}h(t)+J_{21}s(t)+\mu_h b_h(t))\} - h(t)\big].$$

Substituting these expressions into (A11), we obtain the following closed system of equations for $s(t)$ and $h(t)$:

$$\begin{cases} \dfrac{ds(t)}{dt} = w_s\big[\tanh\{\beta(J_{11}s(t)+J_{12}h(t)+\mu_s b_s(t))\} - s(t)\big], \\ \dfrac{dh(t)}{dt} = w_h\big[\tanh\{\beta(J_{22}h(t)+J_{21}s(t)+\mu_h b_h(t))\} - h(t)\big]. \end{cases} \quad (A12)$$

The following condition determine the stationary points of (A12) for $b_s(t) = b_s = \text{const}$, $b_h(t) = b_h = \text{const}$:

$$\begin{cases} s = \tanh(\beta(J_{11}s + J_{12}h + \mu_s b_s)), \\ h = \tanh(\beta(J_{22}h + J_{21}s + \mu_h b_h)), \end{cases} \quad (A13)$$

which coincides with condition (A4) that determines the extrema of the equilibrium distribution function $P_0(s, h)$.

Reverting to notation $\theta = 1/\beta$, we express (A12) as

$$\dot{s} = -w_s s + w_s \tanh\left(\frac{J_{11}s + J_{12}h + \mu_s b_s(t)}{\theta}\right), \quad (A14a)$$

$$\dot{h} = -w_h h + w_h \tanh\left(\frac{J_{21}s + J_{22}h + \mu_h b_h(t)}{\theta}\right). \quad (A14b)$$

## A4. Dynamic equation in the one-component Ising system

Finally, we consider a simpler Ising system that consists of $N$ spins $s_i = \pm 1, i = 1 \ldots N$ and the external magnetic field $H(t)$. The Hamiltonian of this system is equal to

$$\mathcal{H} = -\frac{J_0}{2} \sum_{i \neq k}^{N} s_i s_k - \mu H(t) \sum_{i}^{N} s_i. \qquad (A15)$$

As the Hamiltonian $(A15)$ is a particular case of the Hamiltonian $(A1)$, equations $(A14)$ contain the dynamic equation for average spin in the one-component system: $s(t) = \frac{\langle S \rangle_t}{N} = \frac{S(t)}{N}$. This equation can be derived from $(A14a)$ by setting $J_{12} = 0$ and renaming $\mu_s = \mu$ and $b_s(t) = H(t)$:

$$\dot{s} = -w_s s + w_s \tanh\left(\frac{Js + \mu H(t)}{\theta}\right), \qquad (A16)$$

where $\lim_{N \to \infty}(J_0 N) = J$.

## Appendix B: The reference sentiment level

Here we derive an approximate expression of the reference sentiment level $s_*$, relative to which deviations in sentiment result in a change of market price, as described by the second term in equation (10).

## B1. Equation for $s_*$

As a first step, we take a long-term average of both sides of equation (10) with respect to time to obtain its asymptotic form in which the averaged variables are time-independent:

$$\bar{p} = a_1 \bar{\dot{s}} + a_2 (\bar{s} - s_*), \qquad (B1)$$

where $\bar{p}$ denotes the constant rate of change in market price and $\bar{\dot{s}} = 0$ because $s$ is bounded. Thus, we can rewrite equation (B1) as

$$\bar{\dot{p}} = a_2(\bar{s} - s_*), \quad (B2)$$

where $\bar{s}$ satisfies the similarly averaged equation (4) (with $\bar{\dot{s}} = 0$):

$$\bar{s} = \overline{\tanh(\beta_1 s + \beta_2 H(t))}. \quad (B3)$$

Now, let us consider a situation where the flow of direct information $H(t)$ is on average neutral, i.e. $\overline{H(t)} = 0$. Since according to our market model, it is the flow of direct information that drives market price development, the case of $\bar{H} = 0$ implies that the rate of change in price is on average zero, i.e. $\bar{\dot{p}} = 0$, and thus equation (B2) becomes

$$s_* = \bar{s}. \quad (B4)$$

Therefore $s_*$ equals $\bar{s}$ in the regime where the direct information flow is on average zero, which is the regime described by the potential $U(s)$ for which $c = \beta_2 \bar{H} = 0$ (eq. 6).[37] There is but one caveat to this derivation. Namely that because we have used equation (10) which – being a linear combination of two asymptotic regimes of the actual equation governing price evolution (the true form of which we do not know) – is an approximation, the above relation between $s_*$ and $\bar{s}$ may not be the true relation. However, we expect that this relation holds approximately, as it leads to results that are consistent with observations, as is shown in Section 1.3.2.

We can obtain an approximate solution for $\bar{s}$ from equation (B3) by expanding the hyperbolic tangent in powers of $\beta_1 s + \beta_2 H$ in the neighborhood of $\beta_1 s + \beta_2 H = \overline{\beta_1 s + \beta_2 H} = \beta_1 \bar{s}$ and averaging the resulting series truncated above the quadratic terms. Equation (B3) then takes the approximate form:

---

[37] In the ferromagnetic case ($\beta_1 > 1$), $\bar{s}$ can be nonzero, given non-symmetric initial conditions ($s(0) \neq 0$), finite averaging period and sufficiently small $\sigma_H$.

$$\bar{s} = \tanh(\beta_1 \bar{s}) \left(1 + \sigma^2 (\tanh^2(\beta_1 \bar{s}) - 1)\right), \tag{B5}$$

where $\sigma = \sigma_{\beta_1 s + \beta_2 H}$ is the standard deviation of $\beta_1 s + \beta_2 H$. Using the time series $H(t)$ (Section 1.1) and $s(t)$ (Section 1.2.3), we estimate $\sigma \sim 0.5$, so $\beta_1 s(t) + \beta_2 H(t)$ can be seen to remain reasonably close to the center of expansion $\beta_1 \bar{s}$, which justifies the above approximation for the relevant range of parameter values.

Equation (B5) is applied to determine $\beta_1$ from $s_*$ in the iterative least squares fitting process for the time series $p(t)$ in Section 1.3.2. For this purpose, we apply $\sigma = 0.3$ instead of 0.5. This choice is dictated by the following considerations. Unlike the empirical model (Section 1), the behavior of which is driven primarily by the measured $H(t)$, the theoretical model (Section 2) is sensitive to $\beta_1$. It would therefore be sensible to select $\beta_1$ based on the theoretical model behavior. The estimate of 0.3 results in a more realistic behavior of the theoretical model than 0.5 which leads to higher values of $\beta_1$ than those following from the theoretical model. The choice of 0.3 instead of 0.5 seems acceptable, given the approximate nature of equation (B5) and the inevitable imprecision in measuring and calibrating $H(t)$.

## B2. Perturbative solution for $s_*$

Treating $\sigma^2$ as a small parameter (as follows from the estimates in the preceding section, $\sigma^2 \sim 0.25$ or 0.09), we can construct a perturbative solution to equation (B5). We find that in the leading order the solution is

$$\bar{s} = \tanh(\beta_1 \bar{s}), \tag{B6}$$

i.e. it coincides with the stable equilibrium points of system (4). Consequently, $s_*$ becomes

$$s_* = \bar{s} = s_\pm \text{ for } \beta_1 > 1, \tag{B7a}$$

$$s_* = \bar{s} = 0 \text{ for } \beta_1 \leq 1. \tag{B7b}$$

Solution (B7a) can be understood as follows. If the amplitude of variation in $s$ is so small that a particle cannot leave the well in which it is initially found, then the particle's average position coincides in the leading order with the stable equilibrium point, $s_-$ or $s_+$, in that well. This implies that for very small amplitudes of motion the shape of the well is symmetric around $s_\pm$. Similarly, solution (B7b) means that the particle's average position coincides with the equilibrium point at the origin.

Accounting for the next order correction, $s_*$ can be shown to equal

$$s_* = \bar{s} = s_\pm \left(1 + \sigma^2 \frac{(1-s_\pm^2)}{\beta_1(1-s_\pm^2)-1}\right) \quad \text{for } \beta_1 > 1, \tag{B8a}$$

$$s_* = \bar{s} = 0 \quad \text{for } \beta_1 \leq 1. \tag{B8b}$$

The second term in (B8a) corrects for the asymmetry of the well around $s_\pm$. As shown in Appendix C, $\beta_1(1-s_\pm^2) - 1$ is negative for any $s_-$ and $s_+$, thus the correction term is positive for $s_-$ and negative for $s_+$. Therefore, owing to the asymmetry of the well, the location of $\bar{s}$ is closer to the origin than that of $s_\pm$, which is a result of the outer wall of the well being steeper than the inner wall (Fig. B1). For $\beta_1 \leq 1$, the term (B8b) is zero because in this case the well is symmetric with respect to the equilibrium point at the origin.

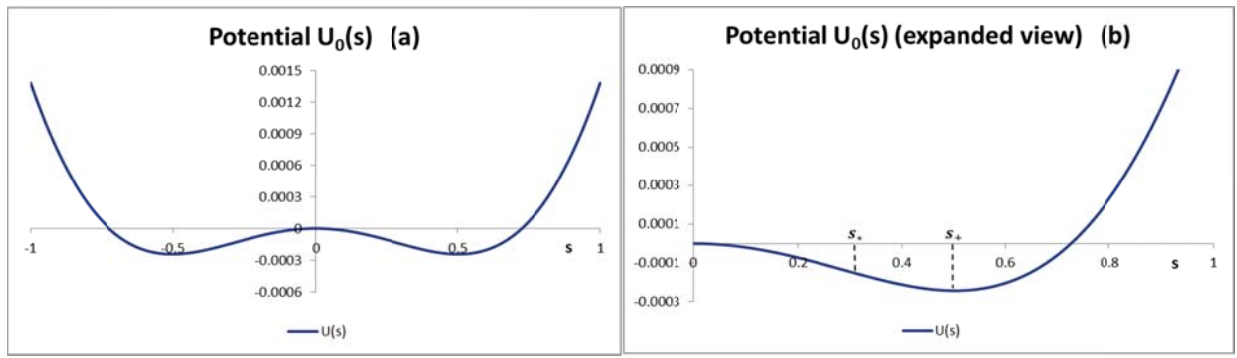

**Figure B1**: Graphic representation of the reference level $s_*$ for $\beta_1 = 1.1$ and $\sigma = 0.3$: $s_+ = 0.503$ (eq. B6) and $s_* = 0.313$ (eq. B8a).

## Appendix C: Analysis of dynamical system

Here we investigate certain properties of the system of equations (13). As equation (13a) is decoupled from (13b,c), the study of system (13) is reduced to the analysis of the two-dimensional nonlinear dynamical system (13b,c), the solutions to which must be substituted into (13a) to solve system (13) completely.

### C1. Equilibrium points

If we consider a situation where the time-dependent force $\kappa \xi(t)$ is absent, equations (13b,c) take the form:

$$\dot{s} = -w_s s + w_s \tanh(\beta_1 s + \beta_2 h), \tag{C1a}$$

$$\dot{h} = -w_h h + w_h \tanh(\gamma \dot{s} + \delta), \tag{C1b}$$

where (as a reminder) $|s| \leq 1, |h| \leq 1$ and the coefficients $\beta_1, \beta_2, \gamma, \delta, w_s$ and $w_h$ are non-negative constants. Solutions to equations (C1) $s(t), h(t)$ for the initial conditions $s(0), h(0)$ determine trajectories of motion corresponding to the velocity field $(\dot{s}, \dot{h})$ given by the r.h.s. of (C1). The motion is bounded because the velocity at the boundaries $s = \pm 1$ and $h = \pm 1$ is directed into the region of motion.

The equilibrium points of system (C1) (where $\dot{s} = \dot{h} = 0$) are given by

$$s^* = \tanh(\beta_1 s^* + \beta_2 h^*), \tag{C2a}$$

$$h^* = \tanh \delta. \tag{C2b}$$

As $\delta$ is given, $h^*$ is known and can be substituted into (C2a) to obtain the equation for $s^*$. It is convenient to write this equation as

$$\beta_2 \tanh \delta = f(\beta_1, s^*) = \frac{1}{2} \ln\left(\frac{1+s^*}{1-s^*}\right) - \beta_1 s^* \qquad (C3)$$

and solve it graphically (Fig. C1). As follows from Figure C1a, the condition that the first derivative of $f(\beta_1, s^*)$ is non-negative at $s^* = 0$, i.e.

$$\left.\frac{\partial f(\beta_1, s^*)}{\partial s^*}\right|_{s^*=0} = 1 - \beta_1 \geq 0,$$

provides for the existence of a single solution to equation (C3). Therefore if $\beta_1 \leq 1$, there exists only one solution $s^* = \tilde{s}_0(\delta)$. The equilibrium solution $\tilde{s}_0(\delta)$ is a single-valued, continuous and odd function of $\delta$, such that $\tilde{s}_0 \to 0$ as $\delta \to 0$. Thus, $\tilde{s}_0$ corresponds to a disordered state – or, in physical terms, a paramagnetic phase – of the system.

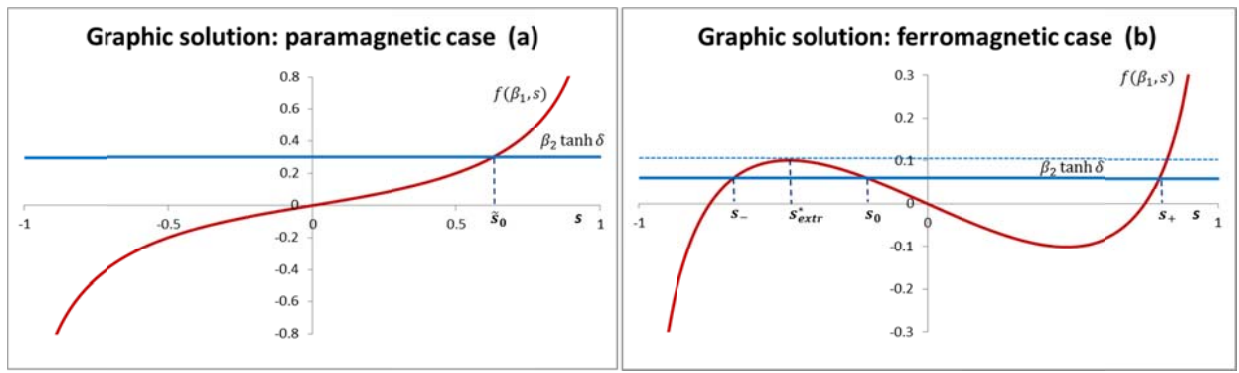

**Figure C1**: Graphic solutions to equation (C3). (a) Paramagnetic case ($\beta_1 \leq 1$). (b) Ferromagnetic case ($\beta_1 > 1$).

If $\beta_1 > 1$, then equation (C3) may have either one or three solutions, depending on the value of $\delta$ (Fig. C1b). When the value of $\delta$ is below critical ($\delta < \delta_c$), equation (C3) has three solutions: $s_-(\delta)$, $s_+(\delta)$ and $s_0(\delta)$, where $s_- < 0$, $s_+ > 0$ and $s_- = -s_+ \neq 0$ in the limit $\delta \to 0$, whereas the third solution $s_0$ is negative for nonzero $\delta$ and approaches zero as $\delta \to 0$. The existence of nonzero solutions in the limit $\delta \to 0$ means that the system may have ordered (ferromagnetic) equilibrium states.

When $\delta \geq \delta_c$, equation (C3) has a single solution $s_+$, as two other solutions merge and vanish at $\delta = \delta_c$. We note that $s_+$ is not sensitive to the bifurcation at $\delta = \delta_c$, that is, $s_+(\delta)$ and its derivatives do not have a singularity at this point. It is not difficult to find $h_c = \tanh \delta_c$. As follows from Figure C1b, $h_c$ is determined by the value of $f(\beta_1, s^*)$ (eq. C3) at its maximum with respect to $s^*$. From the extremum condition, $\frac{\partial f(\beta_1, s^*)}{\partial s^*} = 0$, it follows that

$$s^*_{\text{extr}} = -\sqrt{\frac{\beta_1 - 1}{\beta_1}}, \qquad (C4)$$

where the solution corresponding to the maximum of $f(\beta_1, s^*)$ has been selected. Substituting $s^* = s^*_{\text{extr}}$ into (C3), we obtain

$$h_c = \tanh \delta_c = \frac{1}{\beta_2} f\left(\beta_1, -\sqrt{\frac{\beta_1 - 1}{\beta_1}}\right) = \frac{1}{\beta_2}\left(\frac{1}{2} \ln\left(\frac{1 - \sqrt{\frac{\beta_1 - 1}{\beta_1}}}{1 + \sqrt{\frac{\beta_1 - 1}{\beta_1}}}\right) + \sqrt{\beta_1(\beta_1 - 1)}\right). \qquad (C5)$$

When the system is in the vicinity of the phase transition at $\beta_1 = 1$ from ferromagnetic state into paramagnetic state, then $0 < \beta_1 - 1 \ll 1$, so the above expression for $h_c$ becomes

$$h_c \sim \left(\frac{2}{3\beta_2 \beta_1^{\frac{3}{2}}}\right)(\beta_1 - 1)^{\frac{3}{2}}.$$

To sum up, system (C1) admits two types of bifurcations with respect to the number of equilibrium points. First, there is a phase transition at $\beta_1 = 1$ between the disordered, paramagnetic state with one equilibrium point ($\beta_1 < 1$) and the ordered, ferromagnetic state with three equilibrium points ($\beta_1 > 1$). Second, when the system is in the ferromagnetic state, two equilibrium points merge and vanish at $\delta = \delta_c$, so that only one equilibrium point remains for $\delta > \delta_c$.

## C2. Stability analysis

We proceed to study the stability of system (C1) near the equilibrium points. It is convenient to rescale time as $\tau = \omega_s t$ and rewrite equations (C1a) and (C1b) as

$$\dot{s} \equiv u(s, h) = -s + \tanh(\beta_1 s + \beta_2 h), \qquad (C6a)$$

$$\dot{h} \equiv v(s, h) = -\eta h + \eta \tanh(\bar{\gamma}\dot{s} + \delta), \qquad (C6b)$$

where $\eta = \frac{w_h}{w_s}$ and $\bar{\gamma} = w_s \gamma$.

Linearization of system (C6) in the neighborhood of the equilibrium points $(s^*, h^*)$ leads to solutions $s(\tau)$ and $h(\tau)$ in the form of linear combinations of $\exp(\lambda_- \tau)$ and $\exp(\lambda_+ \tau)$ with eigenvalues $\lambda_\pm$ given by

$$\lambda_\pm = \frac{1}{2}\left(\text{tr}(J) \pm \sqrt{\text{tr}^2(J) - 4\det(J)}\right),$$

where Jacobian $J$, trace $\text{tr}(J)$ and determinant $\det(J)$ are defined as

$$J(s^*, h^*) = \begin{pmatrix} \frac{\partial u}{\partial s} & \frac{\partial u}{\partial h} \\ \frac{\partial v}{\partial s} & \frac{\partial v}{\partial h} \end{pmatrix}\Bigg|_{\substack{s=s^* \\ h=h^*}}, \quad \text{tr}(J) = \frac{\partial u}{\partial s} + \frac{\partial v}{\partial h}, \quad \det(J) = \frac{\partial u}{\partial s}\frac{\partial v}{\partial h} - \frac{\partial u}{\partial h}\frac{\partial v}{\partial s}.$$

Substituting $u(s, h)$ and $v(s, h)$ from (C6) into the expressions for $J$, $\text{tr}(J)$ and $\det(J)$, we obtain

$$J = \begin{pmatrix} -\psi & \frac{1}{\chi}(\varphi + \eta) \\ -\chi\psi & \varphi \end{pmatrix}, \quad \text{tr}(J) = \varphi - \psi, \quad \det(J) = \psi\eta,$$

where

$$\psi = 1 - \beta_1(1 - s^{*2}),$$

$$\chi = \eta\bar{\gamma}\operatorname{sech}^2\delta = w_h\gamma\operatorname{sech}^2\delta,$$

$$\varphi = \beta_2\eta\bar{\gamma}(1 - s^{*2})\operatorname{sech}^2\delta - \eta = \beta_2 w_h\gamma(1 - s^{*2})\operatorname{sech}^2\delta - \frac{w_h}{w_s}.$$

As a result, the characteristic equation for eigenvalues $\lambda_\pm$ becomes

$$\lambda_\pm = \frac{1}{2}\left(\varphi - \psi \pm \sqrt{(\varphi - \psi)^2 - 4\psi\eta}\right). \tag{C7}$$

If the discriminant $D = (\varphi - \psi)^2 - 4\psi\eta$ is non-negative, then $\lambda_\pm$ are real. The eigenvalues may have the same or opposite signs. In the former case the equilibrium point is called a node, in the latter case the equilibrium point is called a saddle. If at least one of the eigenvalues is positive, the equilibrium is unstable because there is an exponentially growing solution to (C6) near the equilibrium point. If $D$ is negative, then $\lambda_\pm$ are complex conjugates, so that there is an oscillatory component to the motion in the vicinity of the equilibrium point, which is called a focus, provided the eigenvalues have nonzero real parts. In this situation, the sign of the real part of the eigenvalues determines stability at the equilibrium. Note that the eigenvalues are dependent on the position of the equilibrium point $s^*$. Therefore, we can expect that each of the three equilibrium points in ferromagnetic state has unique stability properties.

Let us consider the characteristic equation (C7) in detail.

1. If the second term in $D$ is negative, then $\lambda_- < 0$ and $\lambda_+ > 0$, so that the equilibrium point is a (unstable) saddle. This condition is fulfilled when $\beta_1(1 - s^{*2}) > 1$.
2. If $\beta_1(1 - s^{*2}) < 1$, i.e. the second term in $D$ is positive, then the equilibrium is stable (unstable) provided $(\varphi - \psi)$ is negative (positive). This condition can be written as

$$\gamma \lessgtr \frac{1 + \frac{w_s}{w_h}\left(1 - \beta_1(1 - s^{*2})\right)}{w_s \beta_2 (1 - s^{*2})\text{sech}^2 \delta}. \tag{C8}$$

These inequalities determine the values of parameters for which the equilibrium is stable (the sign < above) or unstable (the sign > above).

If simultaneously with (C8) $(\varphi - \psi)^2 > 4\psi\eta$, then $D$ is positive, the eigenvalues are real and of the same sign, so that the equilibrium point is a stable $((\varphi - \psi) < 0)$ or unstable $((\varphi - \psi) > 0)$ node. It is not difficult to show that the corresponding ranges of parameter values are defined by

$$\gamma < \frac{\left(1 - \sqrt{\frac{w_s}{w_h}\left(1 - \beta_1(1 - s^{*2})\right)}\right)^2}{w_s \beta_2 (1 - s^{*2})\text{sech}^2 \delta} \tag{C9}$$

for the stable node and by

$$\gamma > \frac{\left(1 + \sqrt{\frac{w_s}{w_h}\left(1 - \beta_1(1 - s^{*2})\right)}\right)^2}{w_s \beta_2 (1 - s^{*2})\text{sech}^2 \delta} \tag{C10}$$

for the unstable node.

On the other hand, if simultaneously with (C8) $(\varphi - \psi)^2 < 4\psi\eta$, then $D$ is negative, the eigenvalues are complex conjugates, so that the equilibrium point is a stable $((\varphi - \psi) < 0)$ or unstable $((\varphi - \psi) > 0)$ focus. The corresponding ranges of parameter values are determined by

$$\frac{\left(1 - \sqrt{\frac{w_s}{w_h}\left(1 - \beta_1(1 - s^{*2})\right)}\right)^2}{w_s \beta_2 (1 - s^{*2})\text{sech}^2 \delta} < \gamma < \frac{1 + \frac{w_s}{w_h}\left(1 - \beta_1(1 - s^{*2})\right)}{w_s \beta_2 (1 - s^{*2})\text{sech}^2 \delta} \tag{C11}$$

for the stable focus and

$$\frac{1+\frac{w_s}{w_h}\left(1-\beta_1(1-s^{*2})\right)}{w_s\beta_2\left(1-s^{*2}\right)\operatorname{sech}^2\delta}<\gamma<\frac{\left(1+\sqrt{\frac{w_s}{w_h}\left(1-\beta_1(1-s^{*2})\right)}\right)^2}{w_s\beta_2\left(1-s^{*2}\right)\operatorname{sech}^2\delta} \qquad (C12)$$

for the unstable focus.

It follows from (C9-12) that the following sequence of bifurcations always takes place when $\gamma$ increases from zero to infinity: stable node -> stable focus-> unstable focus -> unstable node. Also, equations (C9) and (C10) show that the range of parameter values for the focus-type equilibrium points diminishes with the decreasing ratio $w_s/w_h$ and vanishes in the limit $w_s/w_h \to 0$.

3. In the paramagnetic phase ($\beta_1 < 1$), the condition $\beta_1(1-s^{*2}) < 1$ is satisfied for any $s^*$, therefore the analysis in the paragraph 2 above applies. It means that the equilibrium point $(s^*, h^*) = (\tilde{s}_0, \tanh \delta)$ goes through the above-described series of bifurcations: stable node -> stable focus-> unstable focus -> unstable node.

4. In the ferromagnetic phase ($\beta_1 > 1$), the system may have either (i) three equilibrium points at $s^* = s_-, s_0, s_+$ and $h^* = \tanh \delta$ or (ii) one equilibrium point at $s^* = s_+$ and $h^* = \tanh \delta$, depending on whether $\delta < \delta_c$ or $\delta \geq \delta_c$ (eq. C5), respectively. As follows from Figure C1b, $|s_0| < |s^*_{\text{extr}}|$. Substituting $s^*_{\text{extr}}$ given by equation C4, we obtain that the condition $\beta_1(1-s^{*2}) > 1$ is satisfied for any $s_0$. Conversely, $|s_\pm| > |s^*_{\text{extr}}|$, therefore the condition $\beta_1(1-s^{*2}) < 1$ is satisfied for any $s_-$ and $s_+$. Thus, the equilibrium point corresponding to $s_0$ is always a saddle in accordance with paragraph 1, whereas the equilibrium points corresponding to $s_\pm$ are subject to the sequence of bifurcations stable node -> stable focus-> unstable focus -> unstable node in accordance with paragraph 2.

It is not difficult to show that in the ferromagnetic case with three equilibrium points ($\delta < \delta_c$), each bifurcation from the sequence stable focus-> unstable focus -> unstable node always occurs first at the point for which $s^* = s_-$ and then at the point for which $s^* = s_+$ (Fig. C2a). The bifurcation stable node -> stable focus can occur first at the point with $s^* = s_+$ (Fig. C2b), provided that

$$\frac{\sqrt{1-\beta_1(1-s_+^2)} + \sqrt{\left(\frac{1-s_+^2}{1-s_-^2}\right)\left(1-\beta_1(1-s_-^2)\right)}}{1 + \sqrt{\left(\frac{1-s_+^2}{1-s_-^2}\right)}} < \sqrt{\frac{w_h}{w_s}}$$

$$< \frac{\sqrt{1-\beta_1(1-s_+^2)} - \sqrt{\left(\frac{1-s_+^2}{1-s_-^2}\right)\left(1-\beta_1(1-s_-^2)\right)}}{1 - \sqrt{\left(\frac{1-s_+^2}{1-s_-^2}\right)}}.$$

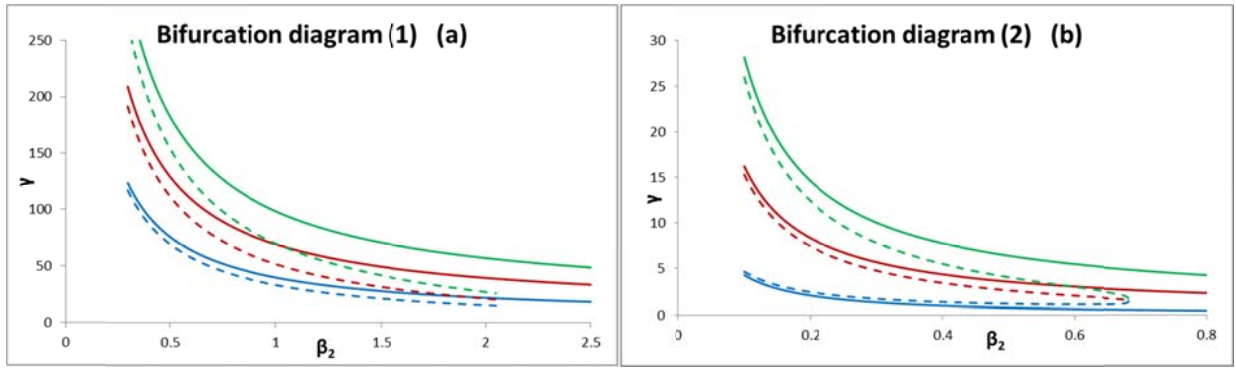

**Figure C2**: The bifurcation diagrams in the $\gamma, \beta_2$-parameter space at the equilibrium points $s_+$ and $s_-$ for $0 < \delta < \delta_c$. Blue, red and green solid lines respectively depict the stable node –> stable focus bifurcation, the stable focus - > unstable focus bifurcation and the unstable focus -> unstable node bifurcation at $s_+$. Blue, red and green dotted lines depict the same sequence of bifurcations at $s_-$. (a) A situation where the bifurcations at $s_+$ occur first ($\beta_1 = 1.3$; $\delta = 0.04$; $w_s = 0.04$; $w_h = 0.4$); (b) A situation where the stable node -> stable focus bifurcation occurs first at $s_-$, while the rest of the sequence remains unchanged ($\beta_1 = 1.1$; $\delta = 0.03$; $w_s = 1$; $w_h = 1$).

## C3. Interpretation as an oscillator

Equation (C6a) can be differentiated with respect to time[38] and, using equation (C6b), expressed as

$$\ddot{s} = \Phi(s, \dot{s}) = -\dot{s} + (1 - (s + \dot{s})^2)(\beta_1 \dot{s} + \beta_2 \eta \tanh(\bar{\gamma}\dot{s} + \delta) + \beta_1 \eta s - \eta \operatorname{arctanh}(s + \dot{s})). \quad (C13)$$

This equation can be interpreted as an equation of motion for a particle of unit mass with the coordinate $s$ and the velocity $\dot{s}$, the motion of which is driven by the applied force $\Phi(s, \dot{s})$. It is instructive to expand $\Phi(s, \dot{s})$ into a Taylor series and write equation (C13) in the following approximate form:

$$\ddot{s} + G(s, \dot{s})\dot{s} + \frac{dU(s)}{ds} = 0, \quad (C14)$$

where $G$ has the meaning of a damping coefficient and is equal to

$$G(s, \dot{s}) = G(s) = (1 - \beta_1 - \beta_2 \eta \bar{\gamma} + \eta) + 2\beta_2 \eta \delta s + (\beta_1 + \beta_2 \eta \bar{\gamma} + 2\beta_1 \eta - 2\eta)s^2 \quad (C15)$$

and the potential $U$ is given by

$$U(s) = -\eta \left( \frac{\beta_1 - 1}{2} s^2 - \frac{\beta_1 - \frac{2}{3}}{4} s^4 + \beta_2 \delta s \right). \quad (C16)$$

To obtain equations (C15) and (C16), the Taylor series of $\Phi(s, \dot{s})$ have been truncated at terms above cubic in $s$, linear in $\dot{s}$ and linear in $\delta$, so that, strictly speaking, this approximation is only valid in the region where $|s| \ll 1$ and $|\dot{s}| \ll 1$, which defines the neighborhood of the equilibrium point at $s^* = s_0$ (and $\dot{s}^* = \tilde{s}_0$) for small $\delta$.[39] However, the expression for the potential $U(s)$, which does not contain the more heavily truncated terms $\sim \dot{s}$, is expected to hold reasonably well also outside that region for the relevant range of parameter values, namely $(\beta_1, \beta_2, \bar{\gamma}) \sim 1$ and $\delta \ll 1$.

---

[38] $d/d\tau$ where $\tau = \omega_s t$.

[39] Recall that both $s$ and $\dot{s}$ are bounded: $|s| \leq 1$ and $|s + \dot{s}| \leq 1$, as follows from equation (C6a).

Equation (C14) is the equation of a damped oscillator that describes the motion of a particle subjected to the restoring force $-dU/ds$ and the damping force $-G\dot{s}$. If damping is not too strong, the motion is expected to be oscillatory. The potential $U(s)$ is useful for visualizing the system's behavior; for instance, the change in the shape of the potential for different $\beta_1$ and $\delta$ helps to explain the occurrence of the bifurcations with respect to the number of equilibrium points (e.g. Fig. C4 below).

The damping coefficient $G$ determines regions where energy is absorbed or supplied. This can be seen by considering the rate of change of the total energy of system (C14):

$$\frac{dE}{dt} = \frac{d}{dt}\left(\frac{1}{2}\dot{s}^2 + U(s)\right) = -G(s)\dot{s}^2.$$

Thus, regions for which $G > 0$ are the regions where energy is extracted from the system (positive or true damping) and regions for which $G < 0$ are the regions where energy is pumped into the system (negative damping). As follows from equation (C15), $G \gtreqless 0$ if

$$\bar{\gamma} \lesseqgtr \frac{(1 - \beta_1 + \eta) + 2\beta_2 \eta\, \delta s + (\beta_1 + 2\beta_1 \eta - 2\eta)s^2}{\beta_2 \eta (1 - s^2)}.$$

This condition, valid near the origin $|s| \ll 1$ for $\delta \ll 1$, shows that damping changes sign from positive to negative with growing $\bar{\gamma}$, provided $\eta > \beta_1 - 1$. In the ferromagnetic symmetric case ($\delta = 0, \beta_1 > 1$), damping becomes negative first at the origin ($s = 0$), so that for the relevant values of parameters, $\beta_1 \gtrsim 1$, $\beta_2 \sim 1$ and $\eta \gg 1$, the critical $\bar{\gamma}$, at which damping changes sign, is given by

$$\bar{\gamma}_c \sim 1, \qquad \text{or equivalently } \gamma_c \sim \frac{1}{w_s} \gg 1.$$

It is worthwhile noting that system (C14)-(C16) contains certain elements of two well-known oscillator models: the Duffing oscillator characterized by a cubic term in restoring force (biquadratic potential) and the van der Pol oscillator characterized by a nonlinear damping that can be negative and a (related) stable limit cycle.

## C4. Phase portraits

Here we discuss the geometry of solutions to system (C1) on the $(s, h)$-plane. Figure C3 shows phase portraits for different values of $\gamma$ in ferromagnetic state that illustrate the sequence of bifurcations at the equilibrium points where $s^* = s_\pm$. Note the emergence of oscillatory motion following the transition stable node -> stable focus and the existence of a large stable limit cycle that encompasses all three equilibrium points during the regimes with the unstable foci and the unstable nodes.

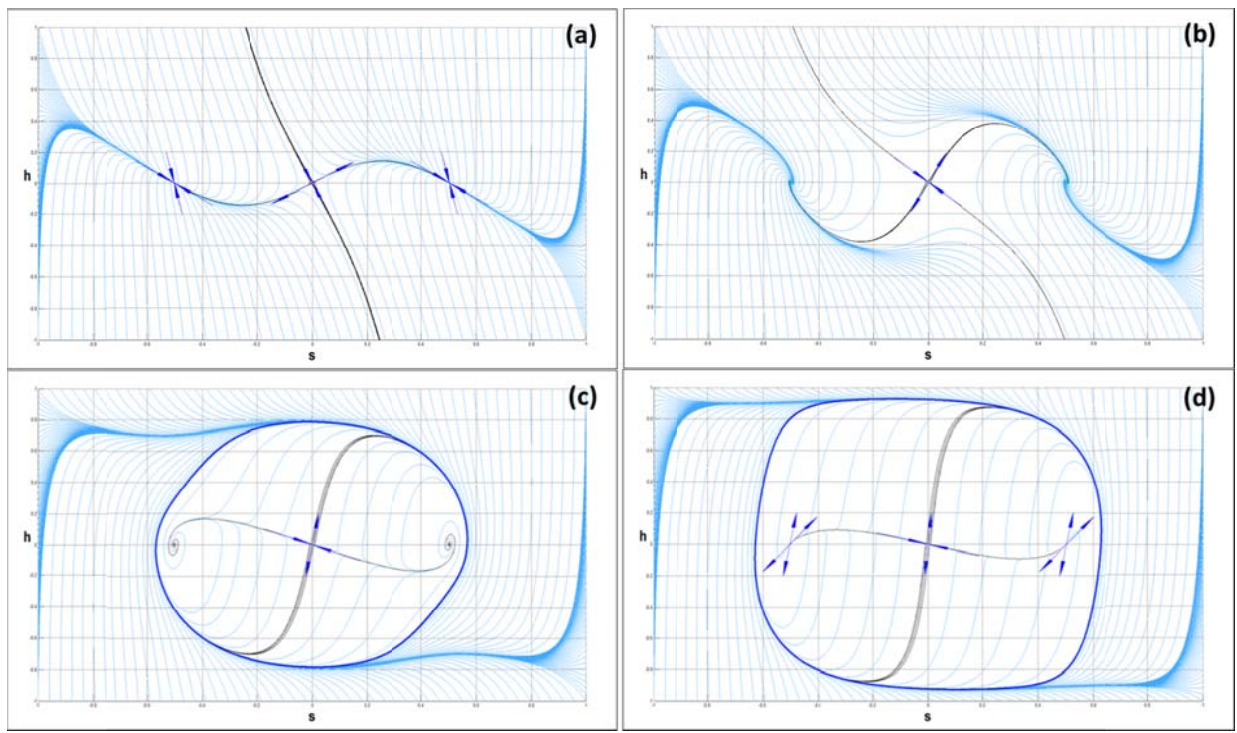

**Figure C3**: Phase portraits at different bifurcation regimes of the equilibrium points $s_\pm$ in the ferromagnetic symmetric case ($\beta_1 = 1.1$; $\delta = 0$): (a) stable nodes ($\bar{\gamma} = 1.6$); (b) stable foci ($\bar{\gamma} =$

2.0); (c) unstable foci ($\bar{\gamma} = 2.6$); (d) unstable nodes ($\bar{\gamma} = 3.4$). Other parameters: $\beta_2 = 0.55$; $w_s = 0.04$; $w_h = 0.4$.

Figure C4 depicts phase trajectories in the ferromagnetic state for $\delta = 0, \delta < \delta_c$ and $\delta > \delta_c$. Note that positive $\delta$ breaks the symmetry of the potential, leading to asynchronous bifurcations in the shallow ($s_-$) and deep ($s_+$) wells. Note also that the equilibrium points $s_0$ and $s_-$ vanish when the extrema of $U(s)$ corresponding to $s_0$ and $s_-$ disappear.

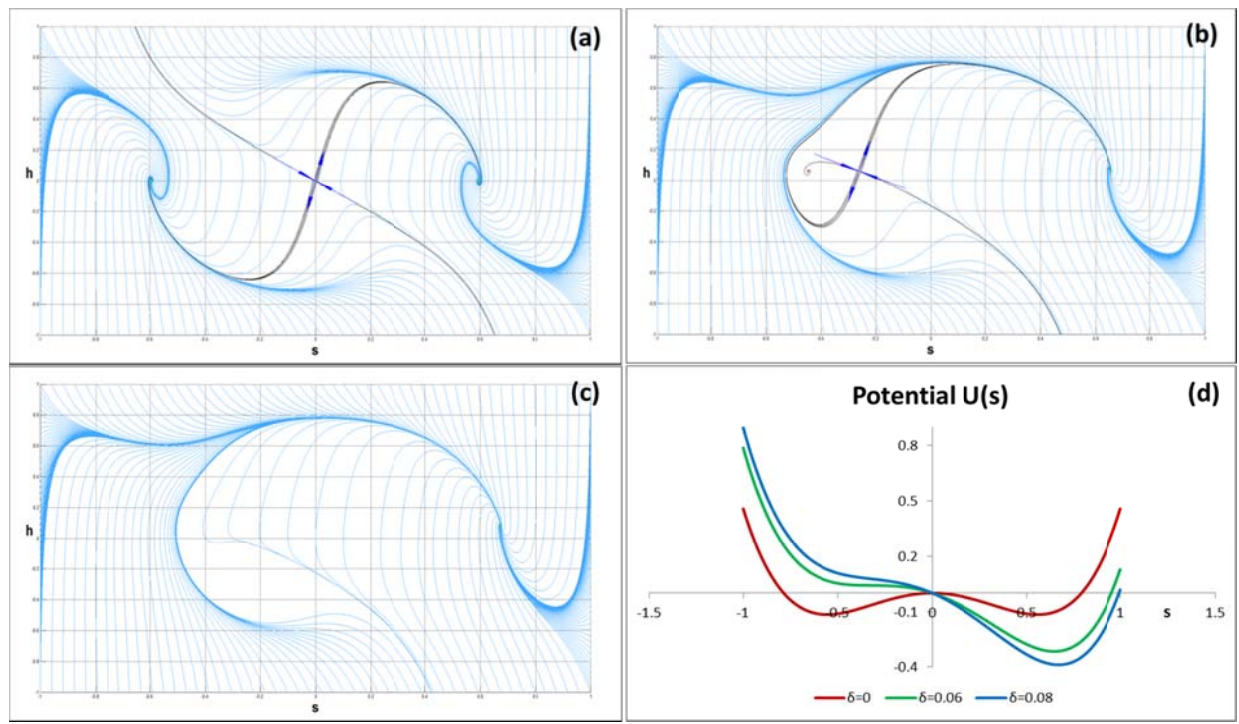

**Figure C4**: Phase portraits for different $\delta$ in the ferromagnetic case ($\beta_1 = 1.15$): (a) $\delta = 0$: two stable foci; (b) $\delta = 0.06$: the stable focus in the shallow well becomes unstable; (c) $\delta = 0.08$: the equilibrium point in the shallow well disappears. (d) The corresponding potentials. Other parameters: $\beta_2 = 0.55$; $w_s = 0.04$; $w_h = 0.4$; $\bar{\gamma} = 2.4$.

Figure C5 shows the stages of formation of a large stable limit cycle in the ferromagnetic state. The stable limit cycle and the enclosed unstable limit cycle emerge as a pair – simultaneously and in the vicinity of each other – in the regime for which the equilibrium points at $s^* = s_\pm$ are stable foci.

As $\gamma$ increases, the limit cycles separate: the stable cycle remains approximately stationary, while the unstable cycle begins to descend toward the equilibrium points and then divides into two small unstable cycles, one around each of the equilibrium points at $s_\pm$, that continue to fall and vanish once they touch the equilibrium points that, at that same moment, bifurcate into unstable foci. The large stable limit cycle continues to exist henceforth.

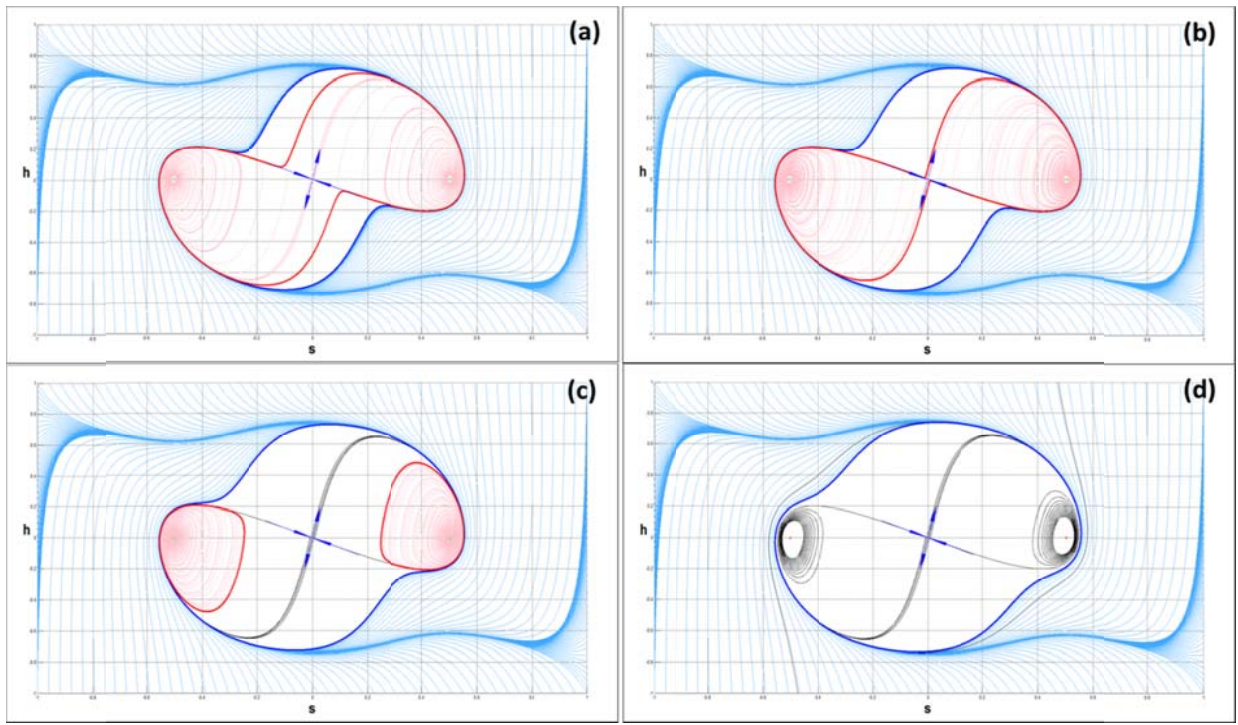

**Figure C5**: Formation of a stable limit cycle in the ferromagnetic symmetric case ($\beta_1 = 1.1$; $\delta = 0$): (a) $\bar{\gamma} = 2.4612$; (b) $\bar{\gamma} = 2.4613$; (c) $\bar{\gamma} = 2.4640$; (d) $\bar{\gamma} = 2.4800$. Other parameters: $\beta_2 = 0.55$; $w_s = 0.04$; $w_h = 0.4$.